\documentclass[twocolumn]{aastex63}
\usepackage{multirow}
\usepackage{lineno}

\accepted{ApJ}

\graphicspath{{./}{figures/}}

\begin{document}
\title{Molecules with ALMA at Planet-forming Scales (MAPS) XIV: Revealing disk substructures in multi-wavelength continuum emission}

\correspondingauthor{Anibal Sierra}
\email{asierra@das.uchile.cl}

\author[0000-0002-5991-8073]{Anibal Sierra}
\affiliation{Departamento de Astronom\'ia, Universidad de Chile, Camino El Observatorio 1515, Las Condes, Santiago, Chile}

\author[0000-0002-1199-9564]{Laura M. P\'erez}
\affiliation{Departamento de Astronom\'ia, Universidad de Chile, Camino El Observatorio 1515, Las Condes, Santiago, Chile}

\author[0000-0002-0661-7517]{Ke Zhang}
\altaffiliation{NASA Hubble Fellow}
\affiliation{Department of Astronomy, University of Wisconsin-Madison, 475 N. Charter St., Madison, WI 53706, USA}
\affiliation{Department of Astronomy, University of Michigan,
323 West Hall, 1085 S. University Avenue, Ann Arbor, MI 48109, USA}

\author[0000-0003-1413-1776]{Charles J. Law}
\affiliation{Center for Astrophysics \textbar\, Harvard \& Smithsonian, 60 Garden St., Cambridge, MA 02138, USA}

\author[0000-0003-4784-3040]{Viviana V. Guzm\'an}
\affil{Instituto de Astrof\'isica, Pontificia Universidad Cat\'olica de Chile, Av. Vicu\~na Mackenna 4860, 7820436 Macul, Santiago, Chile}

\author[0000-0001-8642-1786]{Chunhua Qi} 
\affiliation{Center for Astrophysics \textbar\, Harvard \& Smithsonian, 60 Garden St., Cambridge, MA 02138, USA}

\author[0000-0003-4001-3589]{Arthur D. Bosman}
\affiliation{Department of Astronomy, University of Michigan,
323 West Hall, 1085 S. University Avenue, Ann Arbor, MI 48109, USA}

\author[0000-0001-8798-1347]{Karin I. \"Oberg} 
\affiliation{Center for Astrophysics \textbar\, Harvard \& Smithsonian, 60 Garden St., Cambridge, MA 02138, USA}

\author[0000-0003-2253-2270]{Sean M. Andrews}
\affiliation{Center for Astrophysics \textbar\, Harvard \& Smithsonian, 60 Garden St., Cambridge, MA 02138, USA}

\author[0000-0002-7607-719X]{Feng Long}
\affiliation{Center for Astrophysics \textbar\, Harvard \& Smithsonian, 60 Garden St., Cambridge, MA 02138, USA}

\author[0000-0003-1534-5186]{Richard Teague}
\affiliation{Center for Astrophysics \textbar\, Harvard \& Smithsonian, 60 Garden St., Cambridge, MA 02138, USA}

\author[0000-0003-2014-2121]{Alice S. Booth} 
\affiliation{Leiden Observatory, Leiden University, 2300 RA Leiden, The Netherlands}
\affiliation{School of Physics and Astronomy, University of Leeds, Leeds, UK, LS2 9JT}

\author[0000-0001-6078-786X]{Catherine Walsh}
\affiliation{School of Physics \& Astronomy, University of Leeds, Leeds LS2 9JT, UK}

\author[0000-0003-1526-7587]{David J. Wilner}
\affiliation{Center for Astrophysics \textbar\, Harvard \& Smithsonian, 60 Garden St., Cambridge, MA 02138, USA}

\author[0000-0002-1637-7393]{Fran\c cois M\'enard}
\affiliation{Universit\'e Grenoble Alpes, CNRS, IPAG, F-38000 Grenoble, France}

\author[0000-0002-2700-9676]{Gianni Cataldi}
\affiliation{Division of Science, National Astronomical Observatory of Japan, Osawa 2-21-1, Mitaka, Tokyo 181-8588, Japan}
\affil{Department of Astronomy, Graduate School of Science, The University of Tokyo, 7-3-1 Hongo, Bunkyo-ku, Tokyo 113-0033, Japan}

\author[0000-0002-1483-8811]{Ian Czekala}
\altaffiliation{NASA Hubble Fellowship Program Sagan Fellow}
\affiliation{Department of Astronomy and Astrophysics, 525 Davey Laboratory, The Pennsylvania State University, University Park, PA 16802, USA}
\affiliation{Center for Exoplanets and Habitable Worlds, 525 Davey Laboratory, The Pennsylvania State University, University Park, PA 16802, USA}
\affiliation{Center for Astrostatistics, 525 Davey Laboratory, The Pennsylvania State University, University Park, PA 16802, USA}
\affiliation{Institute for Computational \& Data Sciences, The Pennsylvania State University, University Park, PA 16802, USA}
\affiliation{Department of Astronomy, 501 Campbell Hall, University of California, Berkeley, CA 94720-3411, USA}

\author[0000-0001-7258-770X]{Jaehan Bae}
\altaffiliation{NASA Hubble Fellowship Program Sagan Fellow}
\affil{Earth and Planets Laboratory, Carnegie Institution for Science, 5241 Broad Branch Road NW, Washington, DC 20015, USA}
\affiliation{Department of Astronomy, University of Florida, Gainesville, FL 32611, USA}

\author[0000-0001-6947-6072]{Jane Huang}
\altaffiliation{NASA Hubble Fellowship Program Sagan Fellow}
\affiliation{Department of Astronomy, University of Michigan,
323 West Hall, 1085 S. University Avenue, Ann Arbor, MI 48109, USA}
\affiliation{Center for Astrophysics \textbar\, Harvard \& Smithsonian, 60 Garden St., Cambridge, MA 02138, USA}

\author[0000-0002-8716-0482]{Jennifer B. Bergner}
\affiliation{University of Chicago Department of the Geophysical Sciences, Chicago, IL 60637, USA}
\altaffiliation{NASA Hubble Fellowship Program Sagan Fellow}

\author[0000-0003-1008-1142]{John~D.~Ilee} 
\affil{School of Physics \& Astronomy, University of Leeds, Leeds LS2 9JT, UK}

\author[0000-0002-7695-7605]{Myriam Benisty}
\affiliation{Unidad Mixta Internacional Franco-Chilena de Astronom\'ia, CNRS/INSU UMI 3386, Departamento de Astronom\'ia, Universidad de Chile, Casilla 36-D, Santiago, Chile}
\affiliation{Universit\'{e} Grenoble Alpes, CNRS, IPAG, F-38000 Grenoble, France}

\author[0000-0003-1837-3772]{Romane Le Gal}
\affiliation{Center for Astrophysics \textbar\, Harvard \& Smithsonian, 60 Garden St., Cambridge, MA 02138, USA}
\affiliation{Universit\'{e} Grenoble Alpes, CNRS, IPAG, F-38000 Grenoble, France}
\affiliation{IRAP, Universit\'{e} de Toulouse, CNRS, CNES, UT3, 31400 Toulouse, France}
\affiliation{IRAM, 300 rue de la piscine, F-38406 Saint-Martin d'H\`{e}res, France}

\author[0000-0002-8932-1219]{Ryan A. Loomis}\affiliation{National Radio Astronomy Observatory, 520 Edgemont Rd., Charlottesville, VA 22903, USA}

\author[0000-0002-6034-2892]{Takashi Tsukagoshi} \affiliation{Division of Science, National Astronomical Observatory of Japan, Osawa 2-21-1, Mitaka, Tokyo 181-8588, Japan}

\author[0000-0002-7616-666X]{Yao Liu}
\affiliation{Purple Mountain Observatory \& Key Laboratory for Radio Astronomy, Chinese Academy of Sciences, Nanjing 210023, China}

\author[0000-0003-4099-6941]{Yoshihide Yamato} \affil{Department of Astronomy, Graduate School of Science, The University of Tokyo, 7-3-1 Hongo, Bunkyo-ku, Tokyo 113-0033, Japan}

\author[0000-0003-3283-6884]{Yuri Aikawa}
\affil{Department of Astronomy, Graduate School of Science, The University of Tokyo, 7-3-1 Hongo, Bunkyo-ku, Tokyo 113-0033, Japan}

\begin{abstract}
Constraining dust properties of planet-forming disks via high angular resolution observations is fundamental to understanding how solids are trapped in substructures and how dust growth may be favored or accelerated therein.
We use ALMA dust continuum observations of the Molecules with ALMA at Planet-forming Scales (MAPS) disks and explore a large parameter space to constrain the radial distribution of solid mass and maximum grain size in each disk, including or excluding dust scattering.
In the nonscattering model, the dust surface density and maximum grain size profiles decrease from the inner disks to the outer disks, with local maxima at the bright ring locations, as expected from dust trapping models. The inferred maximum grain sizes from the inner to outer disks decrease from ~1 cm to 1 mm.
For IM\,Lup, HD\,163296, and MWC\,480 in the scattering model, two solutions are compatible with their observed inner disk emission: one solution corresponding to a maximum grain size of a few millimeters (similar to the nonscattering model), and the other corresponding to a few hundred micrometer sizes.
Based on the estimated Toomre parameter, only IM Lup -- which shows a prominent spiral morphology in millimeter dust -- is found to be gravitationally unstable. The estimated maximum Stokes number in all the disks lies between 0.01 and 0.3, and the estimated turbulence parameters in the rings of AS\,209 and HD\,163296 are close to the threshold where dust growth is limited by turbulent fragmentation.
This paper is part of the MAPS special issue of the Astrophysical Journal Supplement.
 \end{abstract}


\section{Introduction} \label{sec:intro}
High angular resolution observations of Class II disks have revealed that most of the large and bright disks have dust morphologies such as gaps and rings \citep[e.g.,][]{Andrews_2016, Isella_2016, Fedele_2017, Avenhaus_2018, Huang_2018, Cieza_2021}, vortices \citep[e.g.,][]{Brown_2009, Andrews_2009, Casassus_2013, Perez_2014, Robert_2020}, and spiral arms \citep[e.g.,][]{Mouillet_2001, Muto_2012, Wagner_2015, Perez_2016, Garufi_2018}.
While the origins of these substructures are still under debate \citep[see discussion in][]{Andrews_2020}, they provide a plausible mechanism for avoiding fast radial migration of dust grains toward the central star \citep{Weidenschilling_1977}. They also prevent the depletion of the raw dust material that can form planets. In addition, these structures may also facilitate planet formation as they concentrate dust grains into narrow spatial regions, which allows for more efficient growth.

Hydrodynamical simulations of protoplanetary disks with rings and gaps \citep[e.g.,][]{Fouchet_2007, Fouchet_2010, Dipierro_2015}, vortices \citep[e.g.,][]{Richard_2013, Surville_2015}, spiral arms \citep[e.g.,][]{Bae_2018}, and analytical models \citep[e.g.,][]{Takeuchi_2002, Pinilla_2012, Lyra_2013, Birnstiel_2013} have shown that gas pressure maxima might act as dust traps, and they could be responsible for the observed substructures at millimeter wavelengths. In these models, the dust mass is enhanced at the pressure maxima and large grains are expected, due to dust size differential trapping \citep[large grains of $\sim$ millimeter and centimeter sizes are more concentrated at the pressure maxima than small grains of $\sim 100 \ \mu$m; e.g.,][]{Ruge_2016}.

The width of the gas pressure bumps is an upper limit to the width of dust rings if the gas pressure maxima can effectively trap dust grains \citep[e.g.,][]{Dullemond_2018, Sierra_2019}. This scenario can be tested using high angular resolution images from molecular line emission and the millimeter dust continuum. The former is difficult, because the optically thin molecules that are needed to trace the underlying gas distributions at high spectral resolution are faint compared with the dust continuum emission. Furthermore, dust rings could be optically thick, complicating the extraction of a reliable gas column density profile.
The width of the gas pressure bumps can also be inferred from the deviation of the gas angular velocity with respect to the Keplerian value \citep[e.g.][]{Rosotti_2020}. Even if the gas angular velocity profile can be estimated from observations, uncertainties in the mass of the central star, limitations in angular resolution, and the  deprojection of the emission-line surface on the midplane can shift the position of the inferred pressure bump and the center of the dust continuum rings.
Dust particles that are radially trapped in rings are also expected to be vertically settled; thus, the local scale height can be used (rather than the radial width of gas pressure bumps) to test dust trapping. Using this method, \citet{Dullemond_2018} found that some of the rings in DSHARP disks \citep{Andrews_2018} are consistent with dust traps.

Dust trapping can also be tested using  multiwavelength dust continuum observations. For example, the dust size differential trapping predicts narrower rings in the dust continuum emission at lower frequencies \citep[e.g.,][]{Pinilla_2015}, or narrower distributions around vortex centers \cite[e.g.][]{Lyra_2013, Sierra_2017}. This behavior occurs because the grain size $a$ traced at a wavelength $\lambda$ follows $a = \lambda/2\pi$ \citep[e.g.,][]{Dalessio_2001}. Thus, if the ring width or vortex area is measured at different frequencies, the presence of dust trapping can be determined. For instance, \cite{Cazzoletti_2018} found a decrease in the vortex area with wavelength, which suggests that dust trapping is active in HD 135344B.   
Moreover, multiwavelength millimeter observations enable a direct determination of dust grain size by fitting the continuum spectrum. In \cite{Carrasco-Gonzalez_2019}, the millimeter continuum spectrum of the HL Tau disk was fitted using ALMA and VLA observations, finding dust grains of millimeter sizes, which tended to be larger in the bright rings and smaller in the gaps. 

Fitting the continuum spectrum also allows estimates of the dust temperature and surface density when sufficient frequency coverage is available. At least four wavelengths are needed to fit the dust surface density, maximum grain size, dust temperature, and compute their covariance matrix. However, such fitting can be performed with a smaller number of wavelengths if additional constraints on the physical properties are available, e.g., dust temperature.
The surface density of solids computed from multiwavelength dust emission is fundamentally important to constrain the dust mass available in disks to form rocky planets. Previous dust mass estimations from many disks seem insufficient to explain the large incidence of massive exoplanets \citep{Greaves_2010, Manara_2018} and the core of giant planets, unless a large fraction of mass is already in large bodies that the millimeter observations are not fully tracing, as proposed by \cite{Najita_2014} in the disks of the Taurus-Auriga region.

\cite{Zhu_2019} provide another possible solution to this problem. They found that optically thick disks can mimic optically thin disks if scattering is properly included in the radiative transfer equation, resulting in the underestimation of the dust mass by a factor from 3 to 30. The emission from an optically thick region can be a factor of 4 smaller than that without scattering if albedo is larger than $\gtrsim 0.6$ \citep{Sierra_2020}. However, scattering can be neglected (even if albedo is high) if the emission is optically thin.

Over the past few decades, the albedo was assumed to be small at millimeter wavelengths, such that scattering effects could be neglected. However, \cite{Miyake_1993} pointed out that scattering opacity could be dominating the dust opacity in disks at millimeter wavelengths.
Since then, many models have included scattering using high albedo \citep[e.g.,][]{Dalessio_2001}. However, nonresolved observations from the optically thick inner disks were not able to observe the effects of scattering. Since ALMA, where the angular resolution was high enough to resolve the inner few tens of astronomical units in disks, many models have used scattering properties to model the inner disks \citep[e.g.,][]{Soon_2017, Liu_2019, Carrasco-Gonzalez_2019, Ueda_2020}.

In this paper, we determine the maximum grain size and dust surface density of five protoplanetary disks, by fitting  spatially resolved multifrequency ALMA continuum observations profiles, and we investigate any evidence of dust trapping in the rings. We compute and compare the results from models with and without scattering in the radiative transfer equation.

This paper is organized as follows. Section \ref{sec:obs} summarizes the dust continuum observations. Section \ref{sec:methodolody} explains how we compute the dust properties, and the results are shown in Section \ref{sec:results}. The optical depth, Toomre parameter, and maximum Stokes number are also estimated and presented.
In Section \ref{sec:discussion}, we discuss our results and how they depend on our assumptions and modeling, as well as highlight notable results for each disk. We present our conclusions in Section \ref{sec:conclusions}.

\section{Observations}\label{sec:obs}

As part of the ALMA Large Program Molecules with ALMA at Planet-forming Scales (MAPS), we obtained continuum observations for the disks around IM Lup, GM Aur, AS 209, HD 163296, and MWC 480 at Bands 3 and 6.
Additional continuum ALMA observations in Bands 4 and 7 were compiled from archival data for AS 209 \citep{Qi_2019} and GM Aur \citep{Huang_2020}. Details on the acquisition of these continuum observations, the calibration of interferometric visibilities, and the self-calibration procedure applied to these data and imaging strategies can be found in \cite{Oberg_2020} and \cite{Czekala_2020}.

For all MAPS observations, we used the self-calibrated visibilities to create one image for each of the four correlator setups for Bands 3 and 6 described in Table 4 of \cite{Oberg_2020}. 
For each correlator setup, the continuum-only spectral window was combined with all other spectral windows where line emission was removed, by discarding channels away from the line center by $\sim15$ km s$^{-1}$, and those near the disk wind in HD 163296 \citep[for details see ][]{Booth_2020}. 
This results in four independent intraband continuum images, corresponding to a central frequency of 94, 106, 226, and 257 GHz or a wavelength of $3.20, 2.84, 1.33$, and $1.17$ mm, respectively. Additionally, the Band 4 and 7 continuum observations for AS\,209 and GM\,Aur were also imaged starting from the self-calibrated visibilities, resulting in two independent continuum images: Bands 4 and 7, corresponding to a central frequency of 145 and 284 GHz or a wavelength of $2.07$ and $1.06$ mm, respectively. 
The continuum images we present here were generated using CASA version 6.1.0 \citep{McMullin_2007} and deconvolved with the \textsc{tclean} task and Briggs weighting with a robust parameter of 0, which achieves a good balance between angular resolution and signal-to-noise ratio (S/N). For all images, the image scales were set to [0, 10, 20, 30, 60] pixels with a cell size of $0.01''$ pixel$^{-1}$. The units of the imaging residuals obtained from CASA are Jy \{dirty beam\}$^{-1}$, while the units of the convolved CLEAN models are Jy \{CLEAN beam\}$^{-1}$; thus, the final CLEANed images are computed after the beam units are corrected. This correction is calculated using the ratio between the volume of the dirty and CLEAN beams, which spans $\sim 0.7-1.1$ for the dust continuum emission of the disks studied. This correction does not modify the total flux of the dust continuum emission because the residuals are small compared with the bright model image.
A full description of this correction (which is called the ``JvM correction" throughout the MAPS papers) can be found in \cite{Czekala_2020}, along with the general imaging strategies.

Table \ref{tab:dust_continuum_prop} summarizes all the relevant imaging properties for each target and each wavelength imaged. The rms noise and peak intensities are directly computed from each map. The rms noise is defined as the standard deviation of the intensity in a box far from the disk, while the peak intensity is the largest intensity within a region that encloses the disk. The peak S/N is the ratio between these two quantities. 

\begin{figure*}[ht!]
    \centering
    \includegraphics[width=0.85\textwidth]{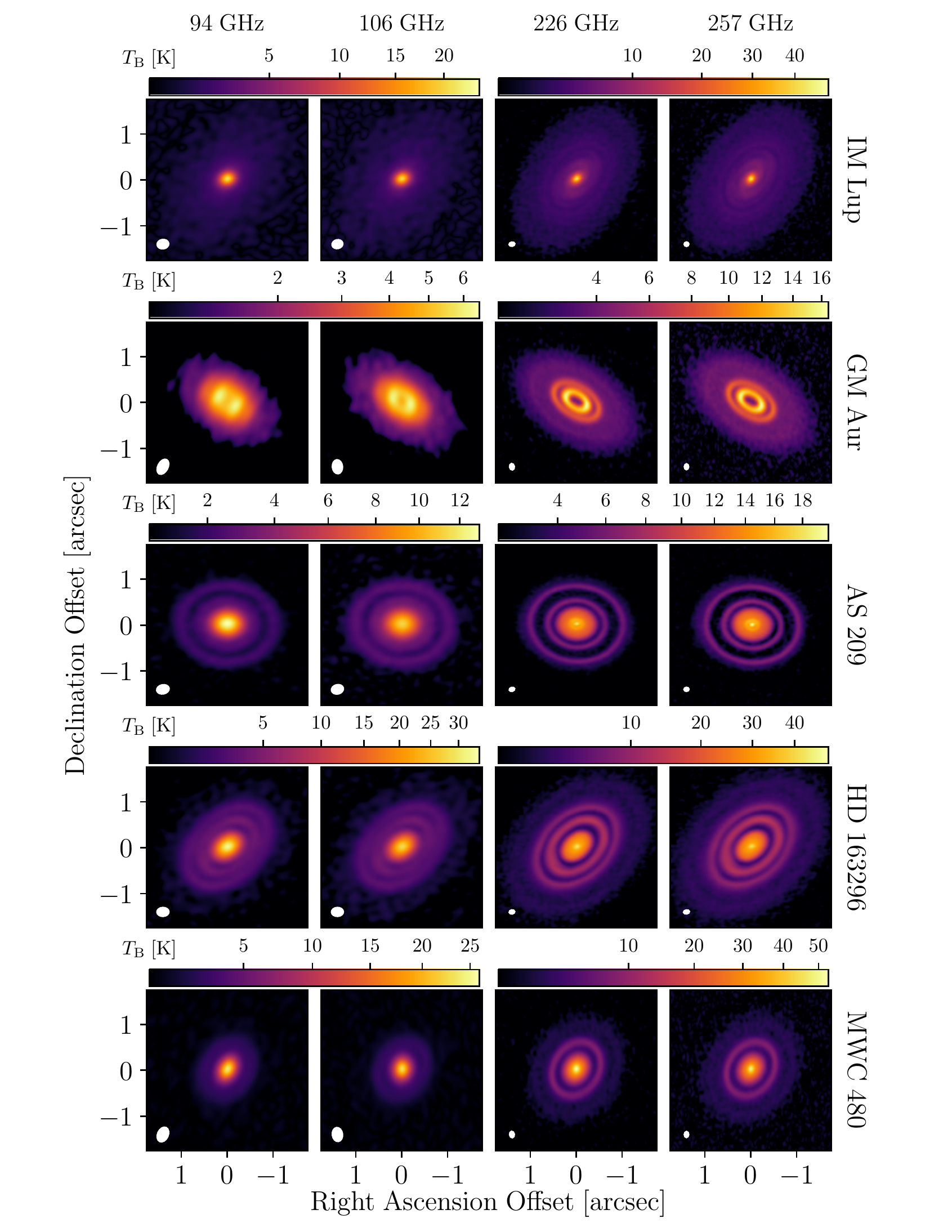}
    \caption{Brightness temperature of the continuum emission from the disks in IM Lup, GM Aur, AS 209, HD 163296, and MWC 480 (ordered by increasing stellar mass; see Table 1 in \citealt{Oberg_2020}), from top to bottom, respectively; and four ALMA Bands at a central frequencies of $94, 106, 226$, and $257$ GHz from left to right. The beam is shown in the lower left corner of each panel. The color bars are stretched such that faint structures in the outer disk can be highlighted.
    Details about each image are described in Table \ref{tab:dust_continuum_prop}.}
    \label{fig:continuum_MAPS}
\end{figure*}

The new images are presented in Figure \ref{fig:continuum_MAPS}, while the Band 4 and 7 images are consistent with those published in \cite{Qi_2019} and \cite{Huang_2020} and thus are not presented. Figure \ref{fig:continuum_MAPS} shows the continuum brightness temperature at each observed wavelength for each disk. The brightness temperature was computed using the full Planck law instead of the Rayleigh-Jeans approximation. Four disks have concentric rings and gaps (AS 209, GM Aur, HD 163296, MWC 480), and one disk (IM Lup) has spiral arms and a gap, only clear in the Band 6 observations. The substructures of these disks in continuum emission are consistent with those observed at similar or higher angular resolution \citep{Long_2018,Huang_2018,Guzman_2018,Isella_2018,Huang_2020}. A full analysis of the radial profiles from the continuum and line emission for each disk can be found in \cite{Law_2020_rad}. In that work, the radial locations of all dust substructures are consistent with previous studies, and in addition, a new ring and gap are also reported in the continuum images: one in MWC 480 and another in IM Lup.

The Band 3 angular resolution and sensitivity are lower than at all other bands, and this is the main limitation of the multiwavelength analysis of these observations (Section \ref{sec:methodolody}). For this reason, all Band 3 visibilities are combined using the \textsc{concat} task in CASA, and then we obtain a Band 3 image with a central frequency of 100 GHz ($\lambda=$ 3.0 mm) for further analysis. The Band 6 data sets are not merged because their angular resolution and sensitivity are higher than the Band 3 images, and they are used to fit the spectral energy distribution (SED) of the disks (Section \ref{sec:methodolody}).

\begin{table*}[ht!]
\centering
\caption{Properties of the millimeter continuum images.}
\begin{tabular}{ccccccccc}
\hline \hline
 \multirow{2}{*}{Source} & ALMA & $\nu$ & \multicolumn{2}{c}{Synthesized Beam} & rms Noise & Peak Intensity & \multirow{2}{*}{Peak S/N} & ALMA \\
        & Band & (GHz) & (mas $\times$ mas; & deg) & ($\mu$Jy beam$^{-1}$) & (mJy beam$^{-1}$) & & Project Code\\
\hline

\multirow{4}{*}{IM Lup} 
 & 3 & 94 & 243 $\times$ 194;& $-$81.5 & 14.1 &  6.44 & 457 & 2018.1.01055.L \\
 & 3 & 106 & 227 $\times$ 182;& $-$77.1 & 14.1 &  6.90 & 489 & 2018.1.01055.L \\    
 & 6 & 226 & 116 $\times$ 81;& $-$80.7  & 19.3 &  15.0 & 777 & 2018.1.01055.L \\
 & 6 & 257 & 91 $\times$ 84;& $-$87.2 & 19.2 &  16.1 & 838 & 2018.1.01055.L \\
\hline

\multirow{6}{*}{GM Aur}
 & 3 & 94 & 341 $\times$ 209;& $-$28.0 & 14.3 & 2.15 & 150 & 2018.1.01055.L \\
 & 3 & 106 & 295 $\times$ 210;&  5.7  & 12.4 & 2.29 & 185 &  2018.1.01055.L\\   
 & 4 & 145 & \ 57 $\times$ 35;& $-$13.8 & 11.3 & 0.48 & 42 & 2017.1.01151.S \\
 & 6 & 226 & 130 $\times$ 94;&  9.1   & 16.7 & 5.83 & 349 & 2018.1.01055.L \\
 & 6 & 257 & 117 $\times$ 83;& $-$1.5   & 21.9 & 5.74 & 262 & 2018.1.01055.L\\
 & 7 & 283 & 270 $\times$ 162;& 0.1   & 57.6 & 22.5 & 390 & 2015.1.00678.S\\
\hline

 \multirow{5}{*}{AS 209}  
 & 3 & 94 & 263 $\times$ 189;& $-$78.3 & 11.6 & 3.94 & 339 & 2018.1.01055.L\\
 & 3 & 106 & 253 $\times$ 193;& $-$74.2 & 13.2 & 3.86 & 294 & 2018.1.01055.L\\    
 & 6 & 226 & 109 $\times$ 77;& $-$86.7  & 15.6 & 4.91 & 314 & 2018.1.01055.L\\
 & 6 & 257 & \ 99 $\times$ 78;& $-$83.1 & 21.9 & 6.12 & 279 & 2018.1.01055.L\\
 & 7 & 283 & 262 $\times$ 180;& $-$73.8 & 54.24 & 32.1 & 592 & 2015.1.00678.S\\
\hline

\multirow{4}{*}{HD 163296} 
& 3 & 94 & 251 $\times$ 184;& $-$88.3 & 11.2 & 10.1 & 902 & 2018.1.01055.L\\
& 3 & 106 & 248 $\times$ 184;& $-$88.1 & 13.2 & 10.7 & 811 & 2018.1.01055.L\\   
& 6 & 226 & 110 $\times$ 80;& $-$80.8  & 17.7 & 14.2 & 802 & 2018.1.01055.L\\
& 6 & 257 & 119 $\times$ 76;& $-$81.5  & 20.3 & 17.6 & 867 & 2018.1.01055.L\\
\hline

\multirow{4}{*}{MWC 480}
 & 3 & 94 & 325 $\times$ 221;& $-$24.5  & 14.6 &  11.3 & 774 & 2018.1.01055.L\\
 & 3 & 106 & 295 $\times$ 212;& 7.5    & 12.4 &  11.8 & 952 & 2018.1.01055.L\\    
 & 6 & 226 & 131 $\times$  93;& 7.5    & 16.9 &  24.5 & 1450 & 2018.1.01055.L\\
 & 6 & 257 & 121 $\times$  85;& $-$1.2   & 22.0 &  24.4 & 1110 & 2018.1.01055.L\\
\hline
\end{tabular}\\
\label{tab:dust_continuum_prop}
\end{table*}

To ensure the highest possible resolution from the data, the radial profiles were independently derived by two methods: first, we reimage all the data using a robust $=-2$ (uniform weighting), in order to increase the angular resolution on the Band 3 image, which sets the limit on the smallest resolution element for this multiwavelength study. The \textsc{restoringbeam} task in CASA is used to obtain a circular beam.
The radial profiles for Bands 4, 6, and 7 are then obtained from convolving all images to the Band 3 beam size using \textsc{imsmooth} in CASA\footnote{\textsc{imsmooth} takes into account the change in the units of the output image by rescaling it by the ratio of the input and output beams} and averaging the emission in concentric ellipses with the inclination and position angle summarized in Table 1 in \cite{Oberg_2020}. The uncertainty of each intensity profile is given by $\Delta I_{\nu} = (\sigma_{I} + {\rm rms}_{\nu}) /\sqrt{n}$, where $\sigma_{I}$ is the standard deviation around the mean intensity, $\rm rms_{\nu}$ is the rms noise in each map, and $n = \Omega_{\rm ring} / \Omega_{\rm beam}$ is the number of beams within a ring of the disk, where $\Omega_{\rm ring}$ and $\Omega_{\rm beam}$ are the solid angles of the ring and beam, respectively. 
Second, we confirm that employing such a robust parameter does not introduce false structures in the radial profiles, as they are consistent with those obtained by fitting the visibilities with a nonparametric model using the software \textsc{Frankenstein} \citep{Jennings_2020}. 
The final image products at the common highest resolution obtained from CLEAN imaging and from \textsc{Frankenstein} are shown in Appendix \ref{app:RadProfiles}.

Figure \ref{fig:intensity_profiles} presents the azimuthally averaged intensity profiles ($I_{\nu}$), which were computed using 10 radial bins per resolution element. The solid lines are the azimuthally averaged intensity for different frequencies (see color legend), and the shaded area is the uncertainty $\Delta I_{\nu}$ of each profile. The dashed lines are the absolute flux calibration uncertainties associated with each band. We set nominal values of 10\% for Band 7 and 5\% for Bands 6, 4, and 3. Note that this error has a magnitude close to $\Delta I_{\nu}$ in the inner disks, while it can be neglected in the outer disks.
The beam size is indicated in the lower left corner of each panel. 
The vertical dashed lines in each panel are the positions of bright rings identified in previous studies (see references in Section \ref{sec:discussion} and the radial structures found by \citealt{Law_2020_rad}) and that can also be distinguished in our data at this resolution.

Figure \ref{fig:spectral_indices} shows the spectral indices ($\alpha = d \log (I_{\nu})/ d \log (\nu)$) of the MAPS observations for each disk between 226 GHz and 100 GHz and between 257 GHz and 100 GHz. The spectral indices were directly computed from the slope of the spectrum at each radial point in the intensity profiles. The shaded areas are the uncertainties in the spectral indices associated with $\Delta I_{\nu}$, while the absolute flux calibration uncertainty is constant at all radii and is represented by two vertical lines in the blank panel.
In general, spectral indices increase with disk radius, from $\alpha \sim 2$ in the inner disk (optically thick emission) to $\alpha \sim 4$ in the outer disk (optically thin), with local minima at the position of the bright rings. This behavior is consistent with previous dust continuum observations, e.g. in the disk around the HL Tau disk \citep{Carrasco-Gonzalez_2019}.

\begin{figure*}[ht!]
    \centering
    \includegraphics[width=\textwidth]{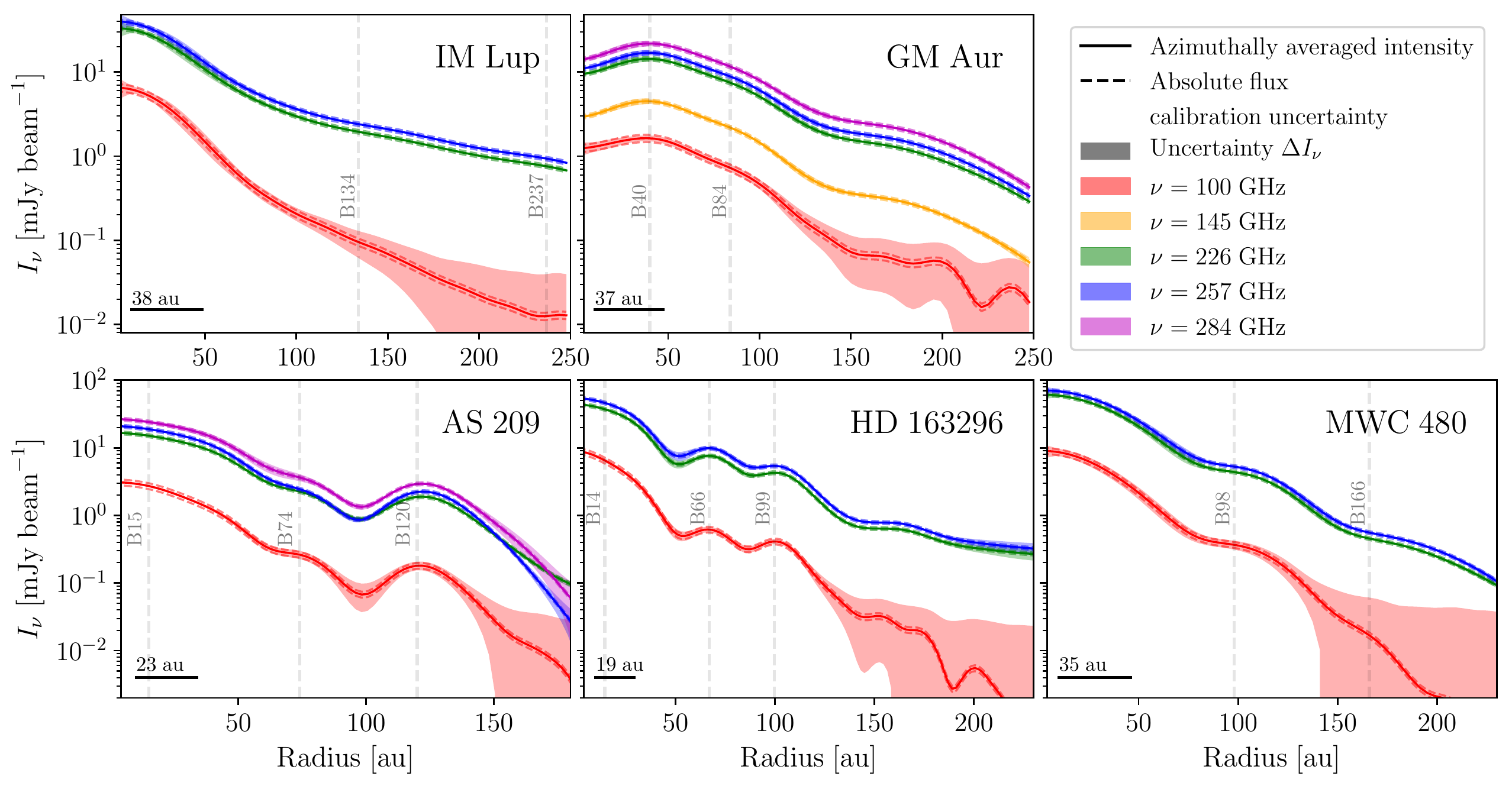}
    \caption{Azimuthally averaged intensity profiles at 100 GHz (red), 145 GHz (orange), 226 GHz (green), 257 GHz (blue), and 284 GHz (magenta). The shaded areas are the uncertainties $\Delta I_{\nu}$ of each profile, and the dashed lines are the absolute flux calibration uncertainty.
    Vertical dashed lines mark the position of bright rings in each disk. The horizontal black line in the lower left corner is the beam size of each disk.}
    \label{fig:intensity_profiles}
\end{figure*}

\begin{figure*}[ht!]
    \centering
    \includegraphics[width=\textwidth]{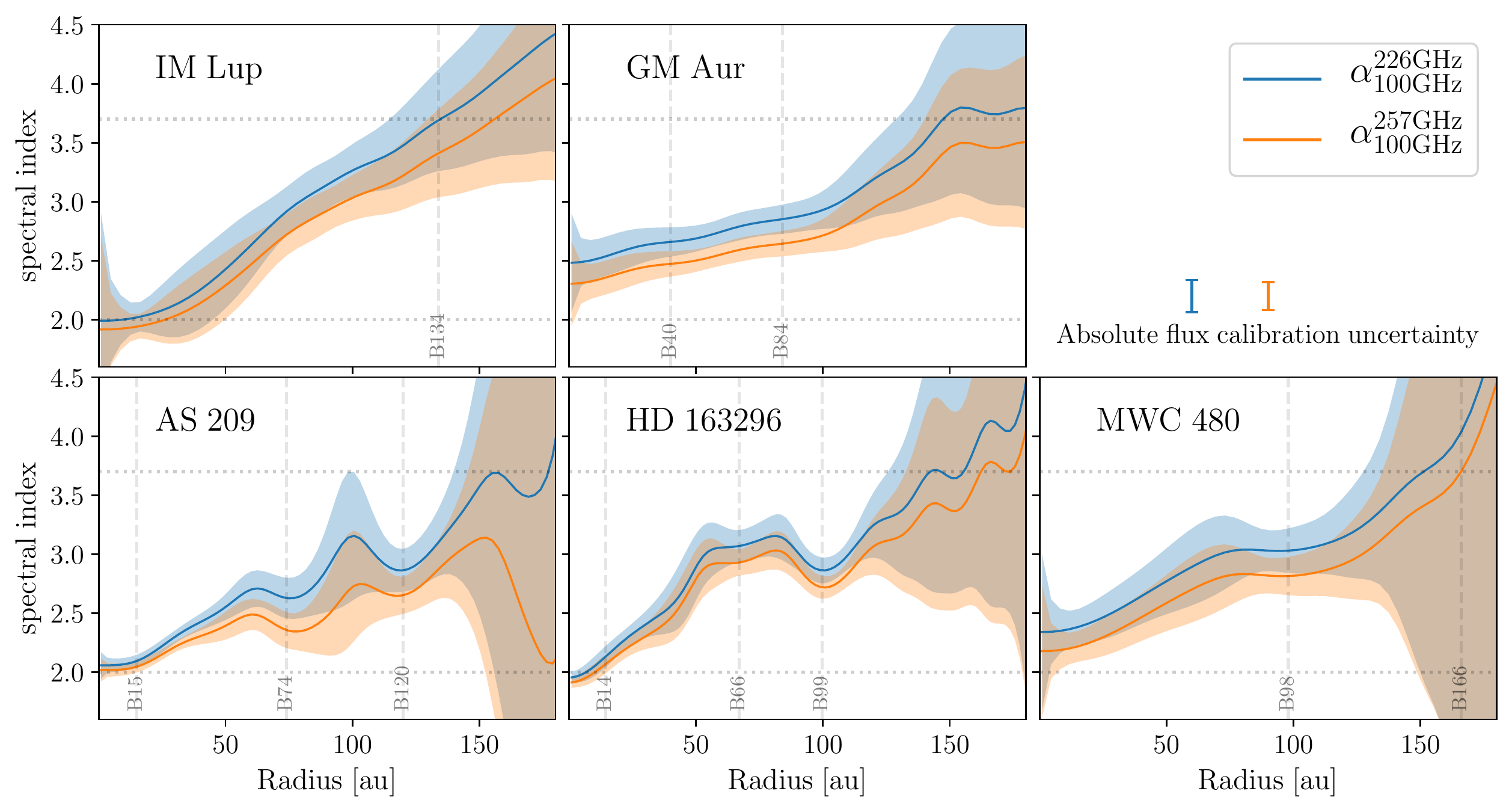}
    \caption{Spectral indices between 100 GHz-226 GHz (blue curves), and 100 GHz-257 GHz (orange curves).
    The shaded areas are the uncertainties of each profile, while the vertical error bars represent the ALMA absolute flux calibration uncertainty.
    Vertical dashed lines mark the position of bright rings in each disk. The horizontal lines indicate $\alpha=3.7$ (typical value in the interstellar medium), and $\alpha=2.0$ (optically thick emission in the Rayleigh-Jeans limit).}
    \label{fig:spectral_indices}
\end{figure*}

\section{Methodology} \label{sec:methodolody}
The dust thermal emission of a plane-parallel slab of the disk is given by
\begin{equation}
    I_{\nu} = B_{\nu}(T_{\rm d}) [1- \exp(-\tau_0 (\nu/\nu_0)^{\beta})],
    \label{eq:Intensity_no_scat}
\end{equation}
where $B_{\nu}(T_{\rm d})$ is the Planck function, $T_d$ is the dust temperature, $\tau_0$ is the optical depth of dust at frequency $\nu_0$, and $\beta$ is the dust opacity spectral index. The temperature, optical depth, and opacity spectral index are parameters that can be fitted to model the multiwavelength dust continuum emission of each disk. However, the optical depth $\tau_{\nu} = \Sigma_{\rm d} \kappa_{\nu}$ at frequency $\nu$ depends on the dust surface density ($\Sigma_{\rm d}$) and the opacity spectrum $\kappa_{\nu}= \kappa_0 (\nu /\nu_0)^{\beta}$, where the opacity coefficient $\kappa_0$ and $\beta$ depend on dust properties such as: dust composition, dust size distribution, maximum grain size, or porosity. In this work we fix the dust composition and the particle size distribution and only consider compact dust spheres (more details about the opacity properties below); thus, the magnitudes of the opacity coefficient and the opacity spectral index are only determined by the maximum grain size ($a_{\rm max}$).

Hence, the dust thermal emission at each radius can be fitted using the dust temperature, dust surface density, and maximum grain size.
If scattering is taken into account, Equation \ref{eq:Intensity_no_scat} is no longer valid. In that case, the solution to the radiative transfer equation for a vertically isothermal slab\footnote{The vertical isothermal slab is a good approximation, as most of the millimeter continuum emission is coming from a thin layer close to the midplane (where the dust scale height is much smaller than the gas scale height), with no abrupt changes in the vertical temperature distribution \citep[e.g.][]{Villenave_2020}.} can be written as \citep{Sierra_2019} 
\begin{equation}
    I_{\nu} = B_{\nu}(T_{\rm d}) [1 - \exp(-\tau_{\nu}/\mu) + \omega_{\nu} F(\tau_{\nu}, \omega_{\nu}) ] ,
    \label{eq:Intensity_scatI}
\end{equation}
where
\begin{eqnarray}
    \nonumber F(\tau_{\nu}, \omega_{\nu}) &=& \frac{1}{\exp(-\sqrt{3}\epsilon_{\nu} \tau_{\nu} ) (\epsilon_{\nu} -1) - (\epsilon_{\nu}+1)} \times \\
    \nonumber && \left[ \frac{1 - \exp(-(\sqrt{3}\epsilon_{\nu}+1/\mu)\tau_{\nu})}{\sqrt{3}\epsilon_{\nu}\mu +1} + \right. \\ 
    && \left. \frac{\exp(-\tau_{\nu}/\mu) - \exp(\sqrt{3}\epsilon_{\nu} \tau_{\nu})}{\sqrt{3}\epsilon_{\nu} \mu -1} \right],\label{eq:Intensity_scatII}
\end{eqnarray}
is the factor that modifies the emergent intensity from the typical nonscattering case, $\epsilon_{\nu} = \sqrt{1-\omega_{\nu}}$, $\mu = \cos(i)$ is the cosine of the disk inclination, and $\tau_{\nu}$ is the total optical depth (absorption + scattering), which can be written in terms of the opacity coefficient and albedo ($\omega_{\nu}$) as $\tau_{\nu} = \Sigma_{\rm d} \kappa_{\nu}/ (1- \omega_{\nu})$. In this work, the albedo is given by the effective albedo \citep{Henyey_1941}, which takes into account the nonisotropic scattering effects via the asymmetry parameter \citep{Birnstiel_2018}. The effective albedo also depends on the maximum grain size.
Equation \ref{eq:Intensity_scatI} reduces to Equation \ref{eq:Intensity_no_scat} when $\omega_{\nu}=0$. 

Equations (\ref{eq:Intensity_scatI}-\ref{eq:Intensity_scatII}) were found by direct integration of the analytical solution of \cite{Miyake_1993} for a vertically isothermal slab, and a similar solution was derived by \cite{Birnstiel_2018} using the Eddington approximation. Both models predict changes in the spectral indices (with respect to the nonscattering case) when scattering is considered. Particularly, scattering can explain spectral indices smaller than the typical value of 2 in the optically thick regime \citep{Liu_2019,Zhu_2019,Sierra_2020}. 

From Equation \ref{eq:Intensity_scatI}, we can constrain the dust properties of each disk by fitting their spatially resolved continuum spectrum from the Band 3 - Band 7 intensity profiles, as was recently done for the HL Tau disk over a similar range of wavelengths \citep{Carrasco-Gonzalez_2019}. We will consider the cases with and without scattering. Even when the former is a more realistic approximation to the radiative transfer effects taking place in protoplanetary disks, there are many previous analyses in disks without scattering, from which we can compare our results.
In order to avoid degeneracy between the dust temperature and dust surface density (see Appendix \ref{app:degeneracy}), the dust temperature of each disk is fixed by the midplane temperature computed in \cite{Zhang_2020}, convolved to the angular resolution of each disk (see Appendix \ref{app:Temperature}). While we assume the same temperature profiles in the scattering and nonscattering cases, in reality they may differ if the albedo is high owing to differing amounts of disk heating \citep[e.g.,][]{Dullemond_2003}.

Additionally, throughout this work, we adopt the dust opacity properties given by \citet{Birnstiel_2018} from the DSHARP collaboration. 
The particle size distribution is assumed to follow $n(a)da \propto a^{-p}da$, with a minimum grain size much smaller ($\sim0.05 \ \mu$m) than the maximum grain size ($\sim$ mm, cm). The effects of the minimum grain size on the opacity properties can be neglected if $p<4$ \citep{Draine_2006}.
Typically, a value of $p=3.5$ is assumed, which was found by \cite{Mathis_1977} in the interstellar medium (ISM). However, the slope of the particle size distribution tends to decrease in protoplanetary disks owing to dust growth \citep{Drkazkowska_2019}.
\cite{Birnstiel_2012} found that this slope changes from $p \sim 3.5$ when the maximum grain size is regulated by fragmentation to $p \sim 2.5$ when it is regulated by drift. 
The magnitude of $p$ determines the opacity properties, in particular, the optically thin spectral index. In our disk sample (Figure \ref{fig:spectral_indices}), the spectral indices are close to $\alpha \sim 3$ at $\sim 100$ au, where one expects optically thin emission.

\begin{figure*}[ht!]
    \centering
    \includegraphics[width=\textwidth]{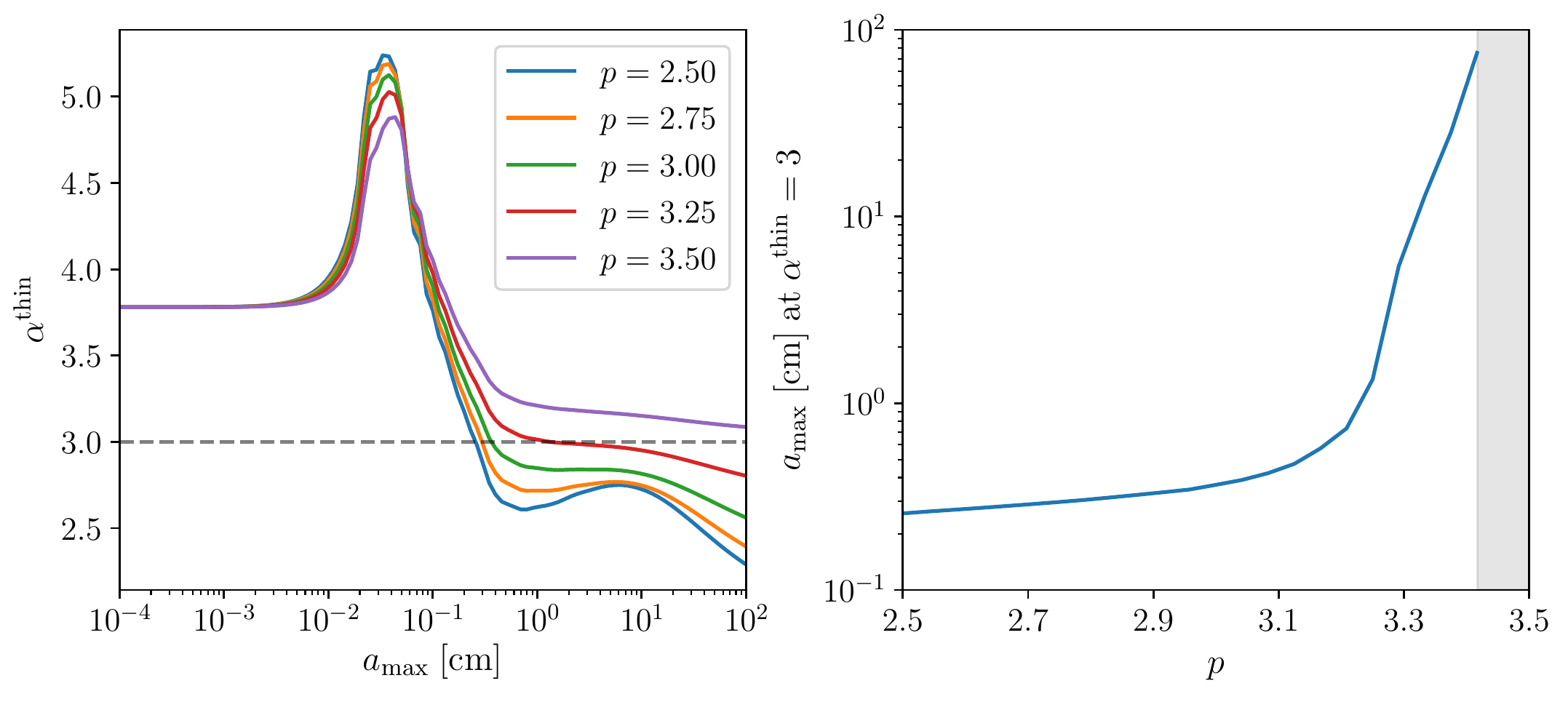}
    \caption{Left panel: optically thin spectral index $\alpha^{\rm thin}$ as a function of the maximum grain size. The curves represent different slopes of the particle size distribution (see color legend). The dashed horizontal line is the reference spectral index in our disk sample at $\sim 100$ au.
    Right panel: maximum grain size as a function of the slope of the particle size distribution when the optically thin spectral index is $\alpha = 3$. The gray shaded area is the region where no solution is found.}
    \label{fig:op_thin_spectral_index}
\end{figure*}

The left panel of Figure \ref{fig:op_thin_spectral_index} shows the optically thin spectral index $\alpha^{\rm thin}$ as a function of the maximum grain size for different slopes $p$. The dashed horizontal line is the reference value of $\alpha = 3$ in the outer disks for our sample. We note that values of $p \gtrsim 3.4$ are not able to explain the reference spectral index with dust grains smaller than 1 m. To visualize this, the right panel of Figure \ref{fig:op_thin_spectral_index} shows the maximum grain size, when the optically thin spectral index is 3, as a function of $p$. This particular maximum grain size rapidly increases for $p \gtrsim 3.2$.
Then, as the magnitude of $p$ cannot be easily constrained, we assume a value of $p=2.5$, which gives a lower limit for the maximum grain size in the disk.

The set of parameters (dust surface density and maximum grain size) that best describes the observations is computed as follows.
At each disk radius, we define the probability of observing the intensities $I_{\nu_1}, I_{\nu_2}, \dots, I_{\nu_n}$ given an intensity model with parameters $a_{\rm max}, \Sigma_{\rm d}$ by the likelihood function 
\begin{equation}
    p(I_{\nu_1}, I_{\nu_2}, \dots, I_{\nu_n} \vert a_{\rm max}, \Sigma_{\rm d}) \propto \exp(-\chi^2/2), 
\end{equation}
where
\begin{equation}
\chi^2 = \sum_n w_{\nu_n} \times \left( \frac{I_{\nu_n} - I_{\nu_n}^{\rm model}}{\epsilon_{\nu_n}} \right)^2,
\label{eq:chi2}
\end{equation}
is the chi-square statistic and $\epsilon_{\nu_n}$ is the total uncertainty of the intensity profiles, given by $\epsilon_{\nu_n}^2 = \Delta I_{\nu_{n}}^2 + (\delta I_{\nu_n})^2$, where $\Delta I_{\nu_{n}}$ is the uncertainty from the intensity profiles (shaved vertical region in Figure \ref{fig:intensity_profiles}), and $\delta$ is the coefficient associated with the uncertainty of the flux calibration, which were described in Section \ref{sec:obs}. The relative weight of each wavelength is given by $w_{\nu_n}$. We choose a relative weight of 2 for Band 3 because two independent data sets were merged to obtain this profile. All the other individual bands have a relative weight of 1.

The explored parameter space is a logarithmically spaced grid between  $10 \ \mu {\rm m} \leq a_{\rm max} \leq 3$ cm and $0.001 \rm \ g \ cm^{-2}<\Sigma_{\rm d}< 10\ g \ cm^{-2}$ (these are sensible ranges in disks), each divided into 200 bins.
The best values for the maximum grain size ($a_{\rm best}$) and the dust surface density ($\Sigma_{\rm best}$) are chosen as those that maximize the likelihood at each radius.
In order to constrain the maximum grain size and dust surface density, we marginalize the joint probability $p(a_{\rm max}, \Sigma_{\rm d})$ and obtain the  probability distributions of these two parameters, $P_{\rm amax}$ and $P_{\rm \Sigma}$.

\begin{figure*}[ht!]
    \includegraphics[width=\textwidth]{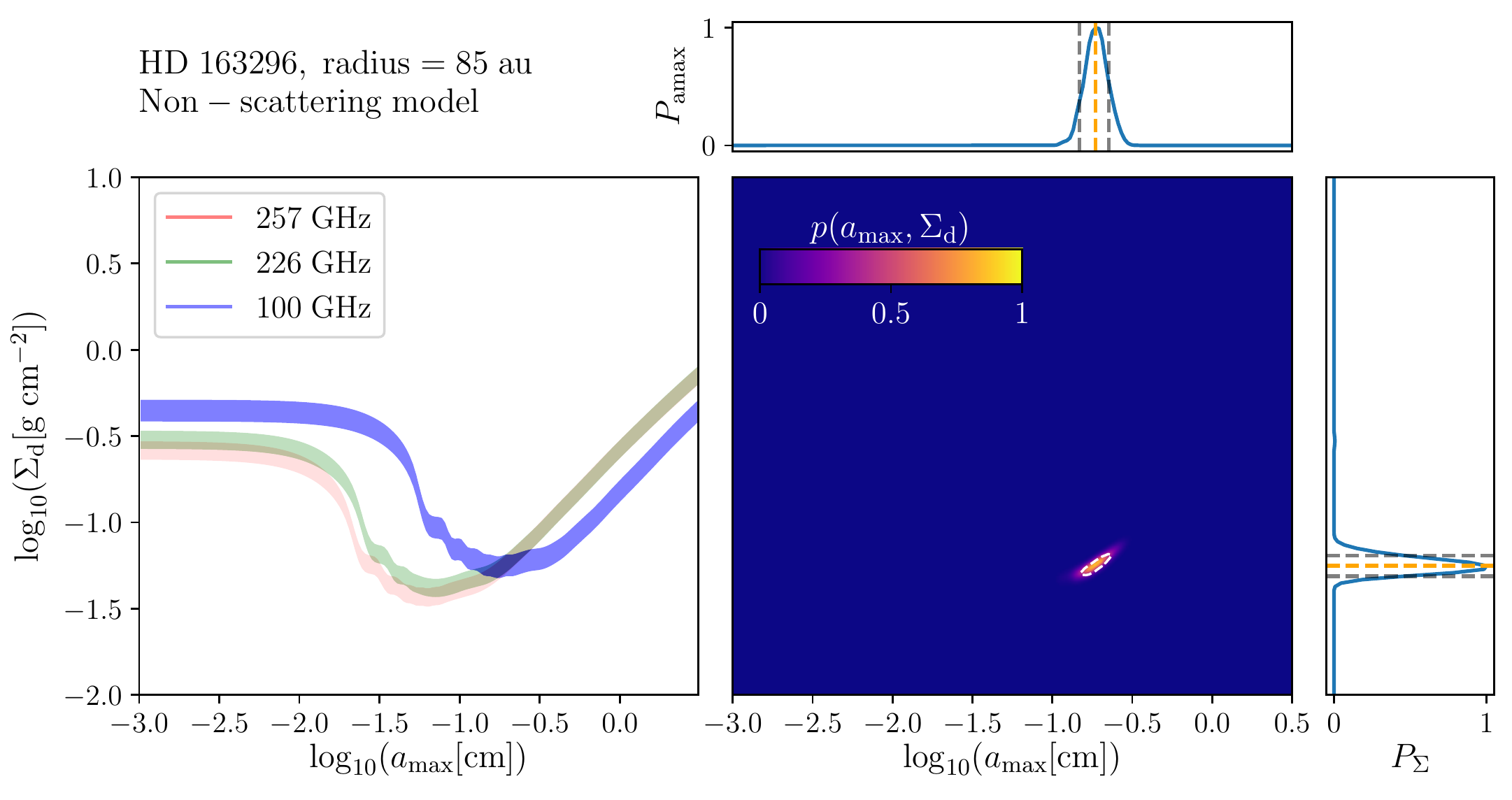}
    \caption{Fitting process for HD 163296 at radius of 85 au in the nonscattering model. Left panel: shaded areas are the regions in parameter space that are able to reproduce the observed emission at each wavelength (see color legend in the left top panel). Right panel: normalized joint probability distribution (color scale); the white dashed line is the 1$\sigma$ contour. Blue lines in the top and right subpanels are the marginal probabilities for $a_{\rm max}$ and $\Sigma_{\rm d}$. Orange dashed lines mark the best-fit values, and gray dashed lines mark the 1$\sigma$ constraints.}
    \label{fig:Fitting_example}
\end{figure*}

As an example of our procedure, Figure \ref{fig:Fitting_example} shows the fitting process for HD 163296 at a radius of 85 au in the nonscattering case. Shaded areas in the left panel are the regions in the parameter space able to reproduce the observed intensity (within the uncertainty) at each wavelength (red, green, and blue for frequencies 257, 226, and 100 GHz, respectively). To fit all observations simultaneously, we need these regions to intersect.
The color scale in the right panel is the normalized joint probability distribution $p(a_{\rm max},\Sigma_{\rm d})$, and the white dashed line is the 1-sigma contour around the maximum probability. Marginal normalized probabilities, $P_{\rm amax}$ and $P_{\rm \Sigma}$, are shown on the top and right of this panel, with orange dashed lines marking the best value and  gray dashed lines marking the 1-sigma range. Repeating this procedure at all sampled radii allows us to infer $a_{\rm max}$ and $\Sigma_{\rm d}$ for the disks studied.

\begin{figure*}[ht!]
    \centering
    \includegraphics[width=\textwidth]{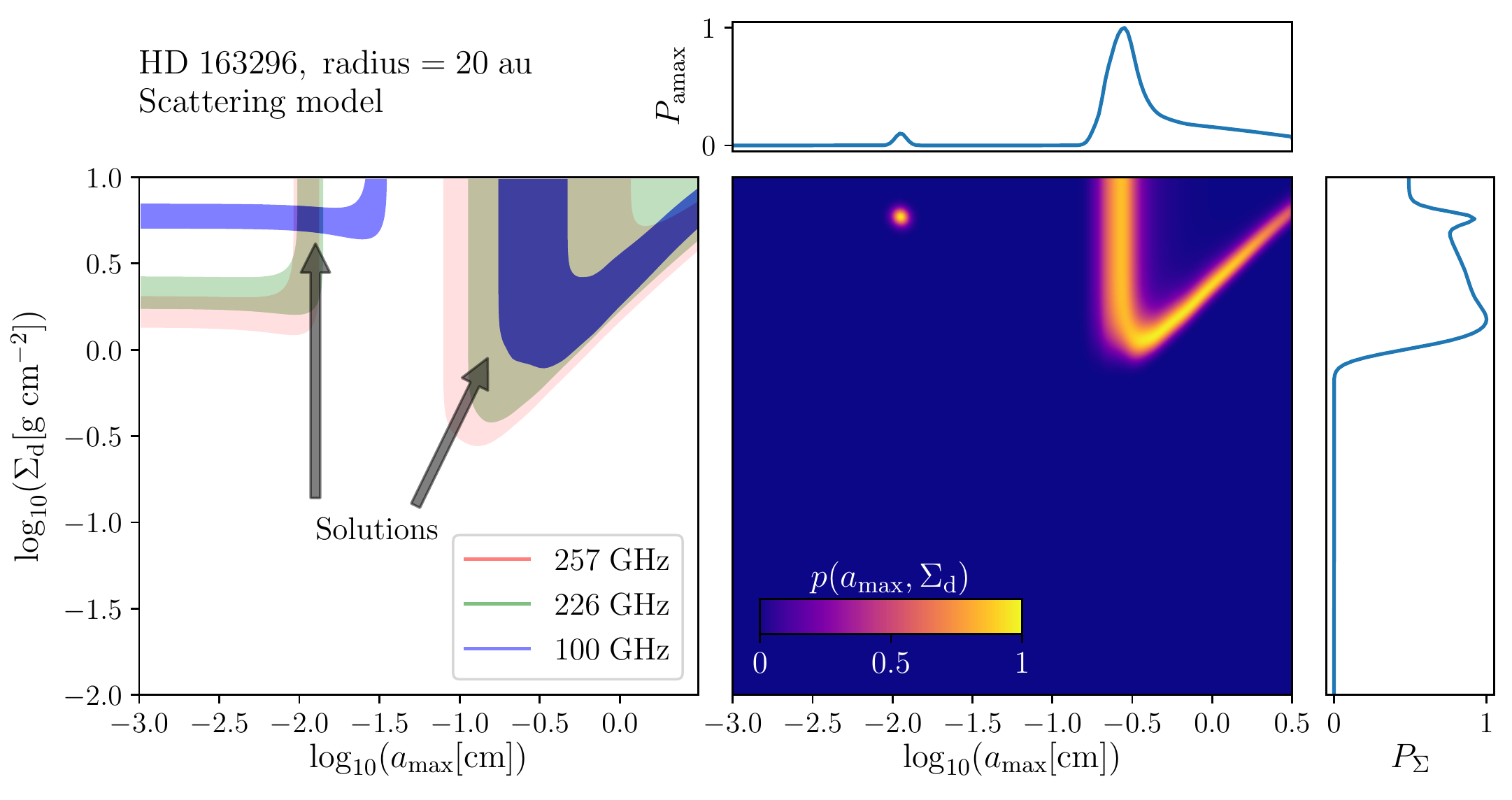}
    \caption{Fitting process for HD 163296 at a radius of 20 au in the scattering model. Left panel: shaded areas are the regions in parameter space that are able to reproduce the observed emission at each wavelength (see color legend in the lower right corner). Right panel: normalized joint probability distribution (color scale). Blue lines in the top and right subpanels are the marginal probabilities for $a_{\rm max}$ and $\Sigma_{\rm d}$.}
    \label{fig:Fit_example_scattering}
\end{figure*}

As an example of our procedure for the model that includes scattering, Figure \ref{fig:Fit_example_scattering} shows the fitting process at 20 au for HD 163296 in the scattering model.
In this case, two different solutions agree with the observations, with $a_{\rm max}$ being either millimeter-sized grains or few-hundred-micron grain sizes. We note that no such degeneracy is present if scattering is neglected.
In the scattering case, this degeneracy appears in the inner disks (the radial extent where this solution is valid for each disk is discussed below) because scattering only modifies the spectral indices when the absorption opacity is optically thick \citep[e.g.,][]{Zhu_2019, Sierra_2020}.
In Figure \ref{fig:Fit_example_scattering}, the left panel shows the regions that are consistent with the emergent intensities of each wavelength within the uncertainties. Dust grains close to $100 \ \mu$m match with the three wavelengths. Additionally, there is a region of large grains ($\gtrsim 1$ mm) that can also fit the data. The right panel is the normalized probability, while the subpanels around this panel are the normalized marginal probabilities ($P_{\rm max}$, $P_{\Sigma}$).

The small-grain and large-grain solutions provide a good fit to the observations, and we are not able to distinguish which one is better. Observations at longer wavelengths (e.g. $\lambda=7$ mm) are needed to dismiss one of the solutions. For that reason, in the scattering case we split the probabilities into two regimes: small grains ($a_{\rm max} < 300 \ \mu \rm m$), and large grains ($a_{\rm max} > 300 \ \mu \rm m$), and we present both results computing the 1-sigma constraints as in the nonscattering case.

\section{Results} \label{sec:results}
Using the radial profiles in Figure \ref{fig:intensity_profiles} and the methodology described in the previous section, the dust surface density and maximum grain size profiles for all the disks are computed and presented here. 
We constrain the radial dust properties for each disk from half of the beam size to the radial position where the enclosed flux at Band 3 (100 GHz) is 95\% of the total flux at this frequency. This percentage is computed using the radial profiles from the CLEANed images in Figure \ref{fig:intensity_profiles}. The influence of this outer radius on the inferred dust masses are discussed in Section \ref{sec:discussion}.

The results are presented as follows: First, we focus the results in the nonscattering model (Section \ref{subsec:no-scattering}), where equation \ref{eq:Intensity_no_scat} is used to fit the SED. Then, scattering effects are taken into account (Section \ref{subsec:scattering}), using the general solution in Equations (\ref{eq:Intensity_scatI})-(\ref{eq:Intensity_scatII}).
In both cases, the optical depths at all wavelengths are computed, and dynamical parameters such as the Toomre parameter and the Stokes number of the maximum grain size are estimated. As mentioned at the end of the previous section, in the scattering case we present two solutions in the inner disk, one of them resulting in millimeter grain sizes and the second one corresponding to some-hundred-micrometer grain sizes.
All the radial profiles of each disk are discussed in Section \ref{sec:discussion}.

\subsection{Nonscattering model}\label{subsec:no-scattering}
\begin{figure*}[ht!]
    \centering
    \includegraphics[width=\textwidth]{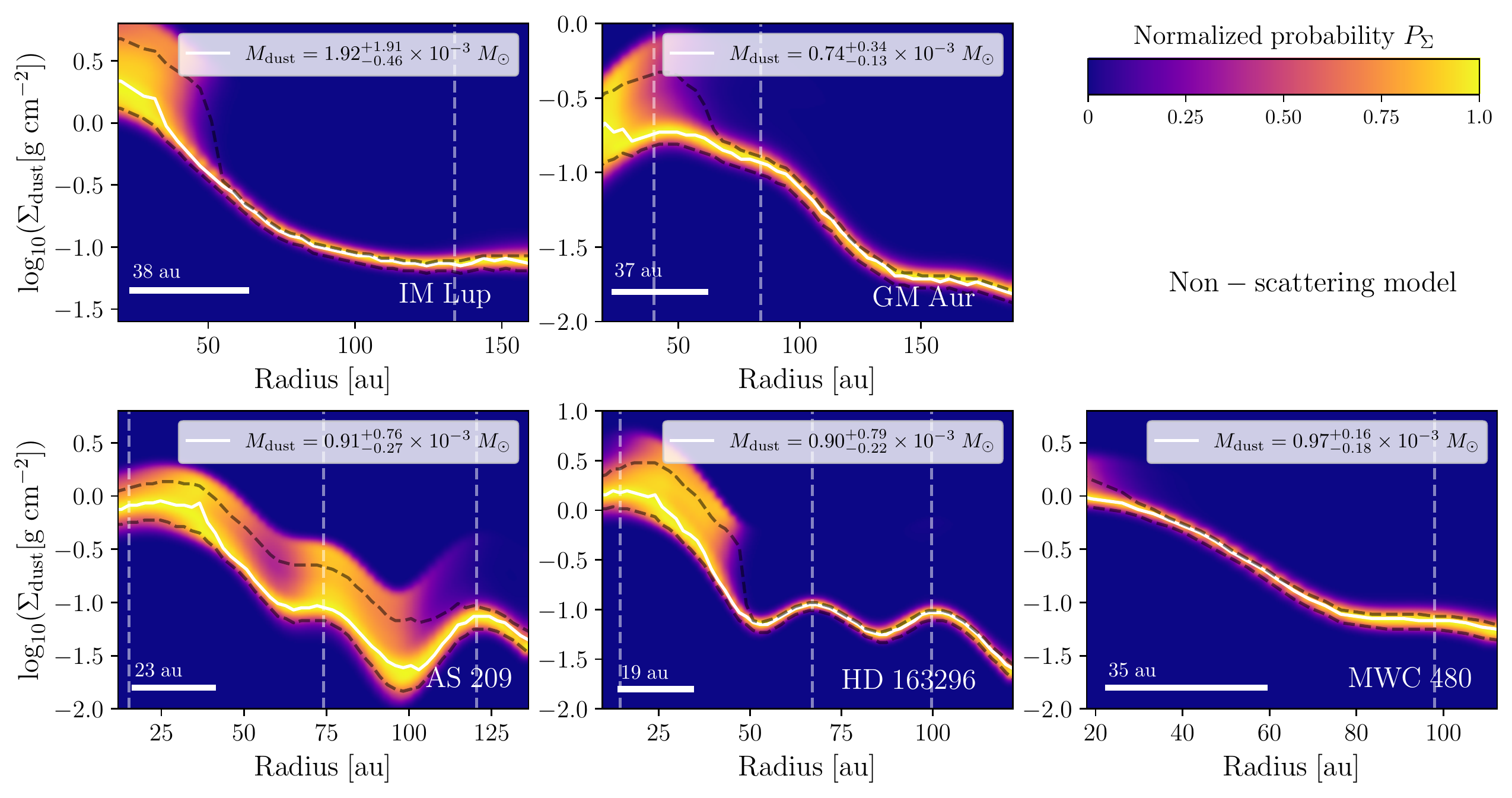}
    \caption{Probability distribution of the dust surface density for all disks in our sample with nonscattering properties. The white solid line is the best-fit dust surface density ($\Sigma_{\rm best}$). The color scale is the marginal probability $P_{\Sigma}$. Dashed black lines are the 1$\sigma$ uncertainties, and the horizontal white line in the lower left corner is the beam size of each disk. Vertical dashed lines mark the position of bright rings. The dust mass in each panel corresponds to the integrated dust surface density within a radius where the enclosed flux at Band 3 is 95\% of the total flux at this band.}
    \label{fig:Results_sigma}
\end{figure*}

\begin{figure*}[ht!]
    \centering
    \includegraphics[width=\textwidth]{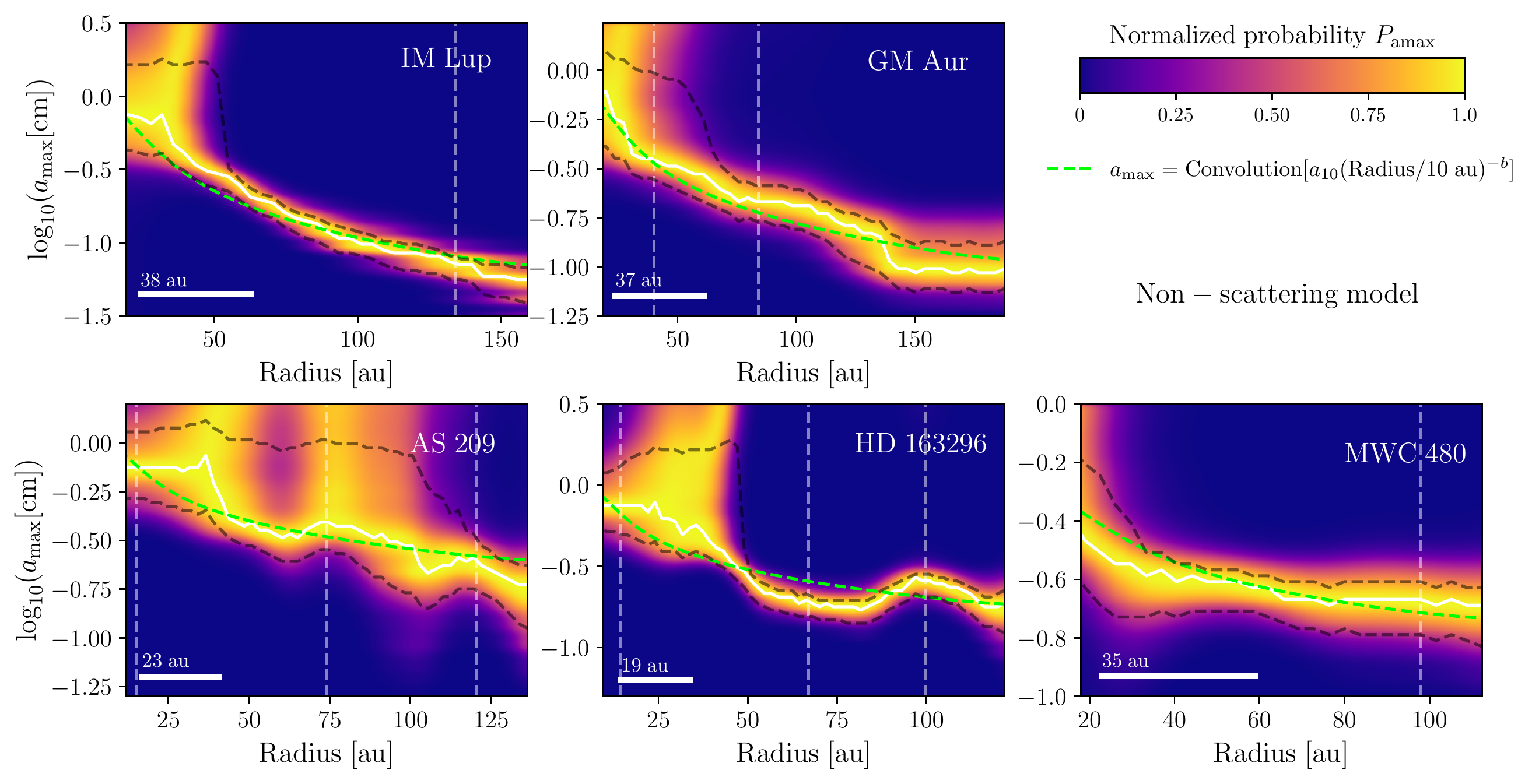}
    \caption{Probability distribution of the maximum grain size for all disks in our sample with nonscattering properties. The white solid line is the best-fit maximum grain size ($a_{\rm best}$). The color scale is the marginal probability $P_{a_{\rm max}}$. Dashed black lines are the 1$\sigma$ uncertainties, and the horizontal white line in the lower left corner is the beam size of each disk. The green dashed line is a power-law fit to the maximum grain size with radius; vertical dashed lines mark the position of the bright rings.}
    \label{fig:Results_amax}
\end{figure*}

Figures \ref{fig:Results_sigma} and \ref{fig:Results_amax} show the probability distributions for the dust surface density and maximum grain size, respectively. In both figures, the white solid lines are the best values ($\Sigma_{\rm best}$, $a_{\rm best}$), while the color scales are the marginal probabilities $P_{\Sigma}$ and $P_{a_{\rm max}}$. The vertical dashed lines mark the positions of the bright rings in each disk that we are able to distinguish in our data \citep[][]{Law_2020_rad}. The spatial resolution (beam size) is shown in the lower left corner of each panel. The black dashed lines in Figures \ref{fig:Results_sigma} and \ref{fig:Results_amax} are the 1
$\sigma$ levels.
The total dust mass for each disk is shown in the upper right legend of each panel.
The dust mass uncertainty is computed from the 1$\sigma$ level of each curve. In all cases, the dust surface density and maximum grain size tend to decrease with the disk radius, with local maxima in the rings. The properties of each disk are discussed in Sections \ref{subsec:IMLup} - \ref{subsec:MWC480}. 

In Figure \ref{fig:Results_amax}, the green dashed line is a deconvolved power law fit to the maximum grain size, i.e. we find a power-law such that, after convolution, fits the maximum grain size radial profiles ($a_{\rm max} = {\rm Convolution}[ a_{10} ({\rm Radius}/10 \ { \rm au})^{-b}]$).
The free parameters of this power law are the maximum grain size at 10 au ($a_{10}$), and the slope $b$ of the power law. These values and the dust masses of each disk are summarized in Table \ref{tab:Dust_inf}.


\begin{table*}[ht!]
    \centering
    \caption{Inferred dust properties in the Noncattering and scattering models with large grains ($a_{\rm max} > 300 \ \mu {m}$).}
    \begin{tabular}{c|ccc|ccc}
    \hline \hline
        \multirow{3}{*}{Disk} & \multicolumn{3}{c|}{Nonscattering model} & \multicolumn{3}{c}{Scattering Model} \\
         & Dust Mass & $a_{10}$ & \multirow{2}{*}{$b$} & Dust Mass & $a_{10}$ & \multirow{2}{*}{$b$}  \\
             & ($\times 10^{-3} M_{\sun}$) & (cm) & & ($\times 10^{-3} M_{\sun}$) & (cm) & \\  
        \hline
        IM Lup    & $1.92^{+1.91}_{-0.46}$ & $1.03 \pm 0.88$ & $0.99 \pm 0.04$ & $3.64^{+1.99}_{-1.41}$ & $1.15 \pm 0.08$ & $1.06 \pm 0.04$ \\
        GM Aur    & $0.74^{+0.34}_{-0.13}$ & $0.82 \pm 0.05$ & $0.70 \pm 0.03$ & $0.73^{+0.32}_{-0.13}$ & $0.82 \pm 0.05$ & $0.70 \pm 0.03$ \\
        AS 209    & $0.91^{+0.76}_{-0.27}$ & $0.83 \pm 0.05$ & $0.47 \pm 0.03$ & $0.75^{+0.72}_{-0.21}$ & $0.79 \pm 0.04$ & $0.47 \pm 0.03$ \\        
        HD 163296 & $0.90^{+0.79}_{-0.22}$ & $0.73 \pm 0.05$ & $0.56 \pm 0.04$ & $0.83^{+1.35}_{-0.18}$ & $0.57 \pm 0.03$ & $0.46 \pm 0.03$ \\
        MWC 480   & $0.97^{+0.16}_{-0.18}$ & $0.49 \pm 0.02$ & $0.41 \pm 0.03$ & $1.19^{+1.77}_{-0.29}$ & $0.31 \pm 0.01$ & $0.21 \pm 0.03$ \\
        \hline
    \end{tabular}
\label{tab:Dust_inf}
\end{table*}

\begin{figure*}[ht!]
    \centering
    \includegraphics[width=\textwidth]{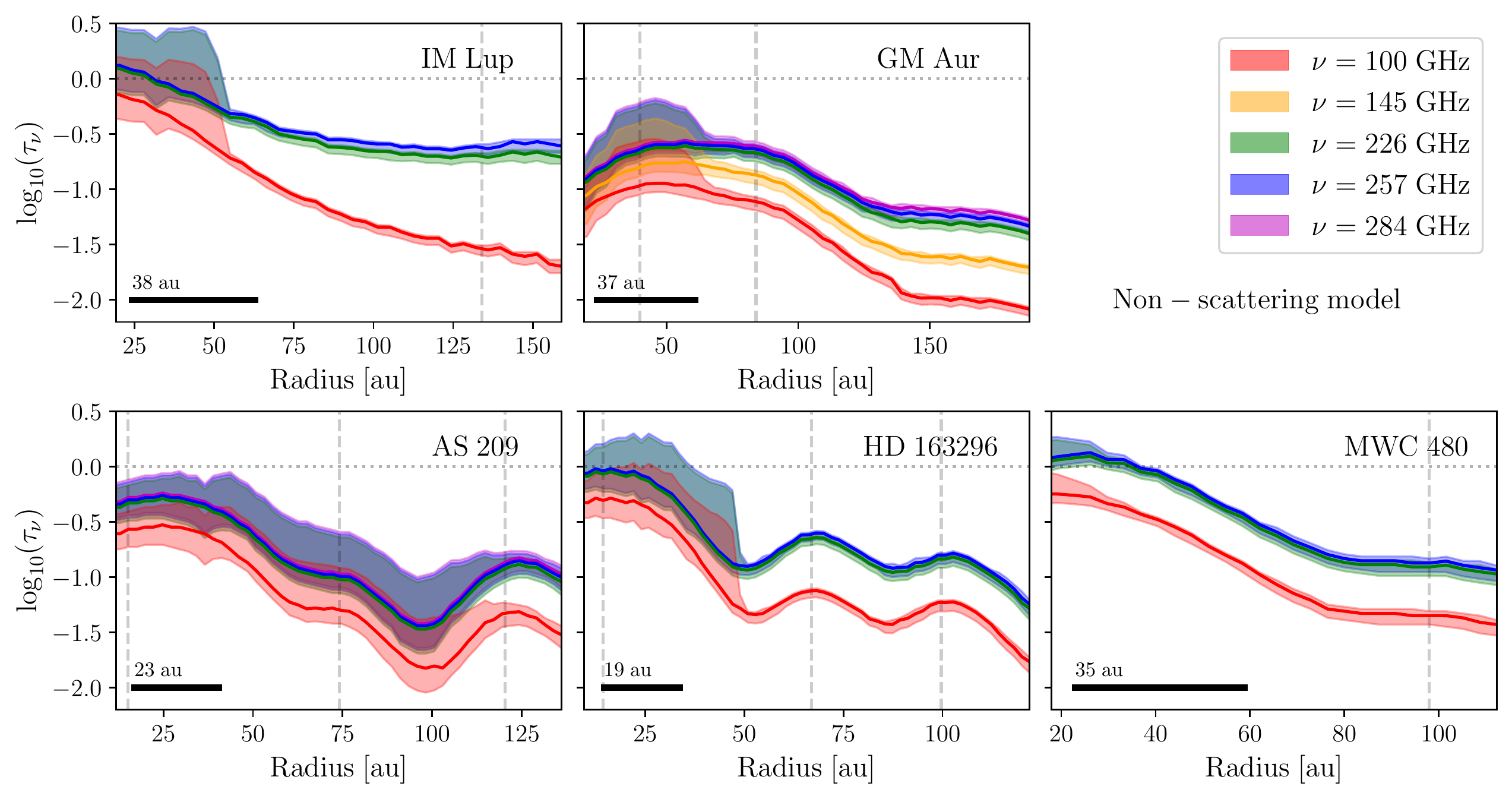}
    \caption{Dust optical depths at 100 GHz (red), 145 GHz (orange), 226 GHz (green), 257 GHz (blue), and 284 GHz (magenta) for the five disks studied in the nonscattering model. The shaded areas are the uncertainties of each profile.
    The horizontal black line in the lower left corner is the beam size of each disk. Vertical dashed lines mark the position of the bright rings, and the horizontal dotted line is a reference value where the optical depth is 1.}
    \label{fig:optical_depth}
\end{figure*}

The optical depths at different wavelengths are computed as $\tau_{\nu} = \Sigma_{\rm d} \kappa_{\nu}$, where the dust opacity coefficient is computed adopting the inferred maximum grain size. Figure \ref{fig:optical_depth} shows the optical depths of the five disks at different wavelengths. Three of the disks (IM Lup, HD 163296, MWC 480) have an optical depth close to or above 1 at Band 6 in the inner disks ($\lesssim 20$ au), but GM Aur and AS 209 have an optical depth $<$ 1 everywhere.

Additionally, the Toomre parameter ($Q$) and the Stokes number ($\rm St (a_{\rm{max}})$) of the maximum grain size can be estimated from the above results. The former quantifies the gravitational disk stability, in particular, if $1<Q<1.7$, the disk is stable to linear perturbations, but second-order perturbations could grow and make the disk gravitationally unstable \citep{Toomre_1964}. If $Q<1$, the disk is also unstable to linear perturbations.

The Stokes number describes the dust dynamics, in particular, it quantifies the coupling between dust and gas \citep{Whipple_1972}. Dust grains with a large Stokes number (${\rm St} \gg 1$) are decoupled from the gas, i.e.\ they are not strongly affected by the gas dynamics, while dust grains with a small Stokes number ($\rm St \ll 1$) are coupled with the gas.
These parameters are defined as
\begin{eqnarray}
\label{eq:Toomre} Q &=& \frac{c_s \Omega}{\pi G \Sigma_{\rm g}}, \\ 
\label{eq:Stokes} \rm{St}&=& \frac{\pi \rho_{\rm m} a}{2 \Sigma_{\rm g}},
\end{eqnarray}
where $c_s$ is the sound speed; $\Omega$ is the disk angular rotation frequency (assumed to be Keplerian); $G$ is the gravitational constant; $\Sigma_{\rm g}$ is the gas surface density;  $\rho_{\rm m} = 1.675$ g cm$^{-3}$ \citep{Birnstiel_2018} is the dust bulk density, computed from the dust composition mix of solids (water ice, silicates, troilite, and refractory organics); and $a$ is the grain size. In particular, we compute the Stokes number associated with the maximum grain size $\rm{St}(a_{\rm max})$. We subsequently refer to this parameter as the maximum Stokes number.
The sound speed is computed from the midplane temperature as
\begin{equation}
    c_s^2 = \frac{k_{\rm B}T}{\mu_m m_{\rm H}},
\end{equation}
with a mean molecular weight $\mu_m = 2.34$, and where $k_B$ and $m_{\rm H}$ are the Boltzmann constant and the hydrogen mass, respectively. 

\begin{figure*}[ht!]
    \centering
    \includegraphics[width=\textwidth]{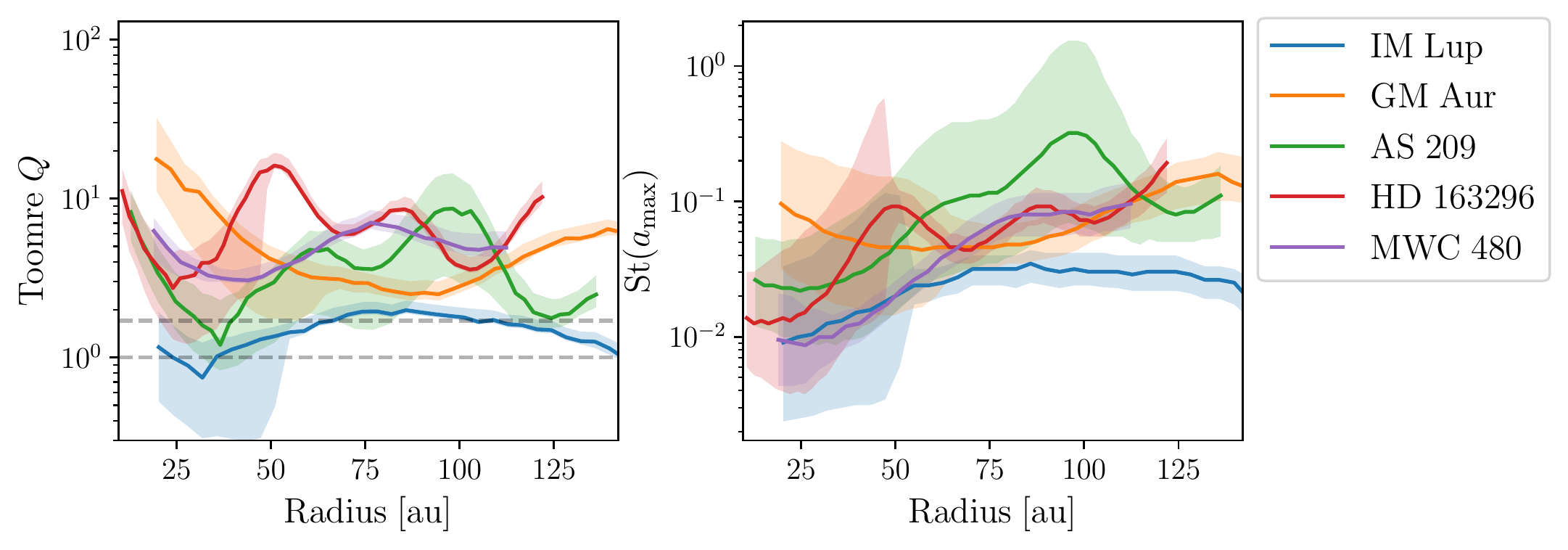}
    \caption{Toomre $Q$ parameter (left) and maximum Stokes number (right) for the five disks (see legend on the left panel) assuming a dust-to-gas mass ratio of 1/100. The horizontal dashed lines in the left panel ($Q <$ 1, 1.7) delimit the regions where the disks become gravitationally unstable to linear and second-order perturbations, respectively. The shadowed areas are the uncertainties of each curve, which were computed from the uncertainty of the dust surface density and maximum grain size.}
    \label{fig:toomre_stokes}
\end{figure*}

The gas surface density is assumed to be 100 times the dust surface density. The influences of these assumptions on the results are discussed in Section \ref{sec:discussion}. Figure \ref{fig:toomre_stokes} shows the Toomre parameter (left panel) and the Stokes number (right panel) of the five disks (see color legend). Most disks appear to be gravitationally stable, except IM Lup, while the Stokes number is below 1 across the sample at all radii (the properties of each disk are discussed in Sections \ref{subsec:IMLup}-\ref{subsec:MWC480}).

\subsection{Scattering model}\label{subsec:scattering}
We now consider the case where scattering is taken into account in the radiative transfer calculations (Equations (\ref{eq:Intensity_scatI})-(\ref{eq:Intensity_scatII})) and recompute the probability distributions. As mentioned in Section \ref{sec:methodolody}, the inner disks have two possible solutions, one with large grains ($a_{\rm max}>300 \ \mu$m) and a second one with small grains ($a_{\rm max}<300 \ \mu$m).

\begin{figure*}[ht!]
    \centering
    \includegraphics[width=\textwidth]{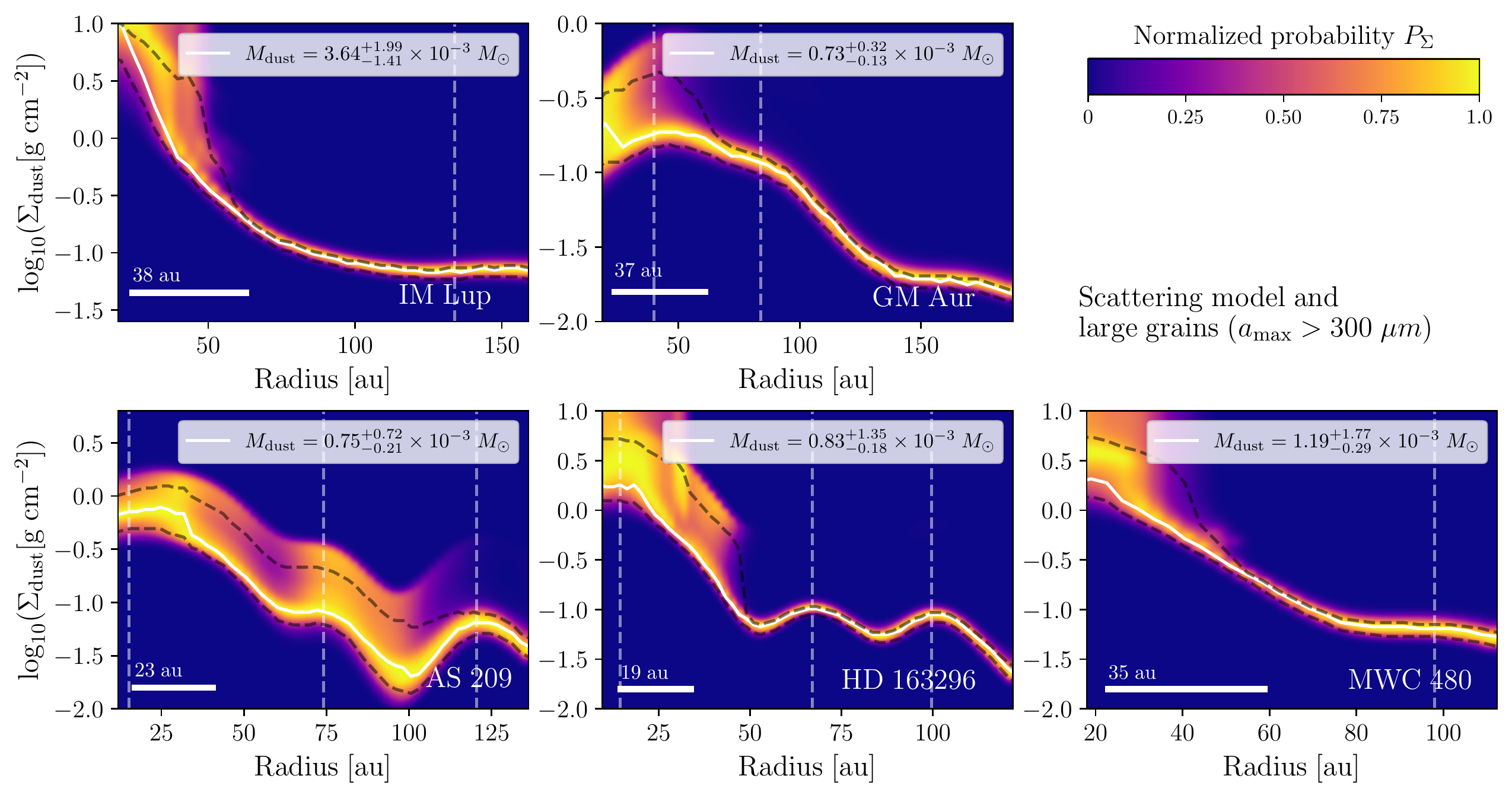}
    \caption{Probability distribution of the dust surface density for all disks in our sample when scattering is taken into account in the radiative transfer model and the large-grain solution ($a_{\rm max} > 300 \ \mu$m) is considered for the inner disk. The white solid line is the best-fit dust surface density ($\Sigma_{\rm best}$). The color scale is the marginal probability $P_{\Sigma}$. Dashed black  lines are the 1$\sigma$ uncertainties, and the horizontal white line in the lower left corner is the beam size of each disk. Vertical dashed lines mark the position of the bright rings. The dust mass in each panel corresponds to the integrated dust surface density within a radius where the enclosed flux at Band 3 is 95\% of the total flux at this band.}
    \label{fig:Results_sigma_Scat}
\end{figure*}

\begin{figure*}[ht!]
    \centering
    \includegraphics[width=\textwidth]{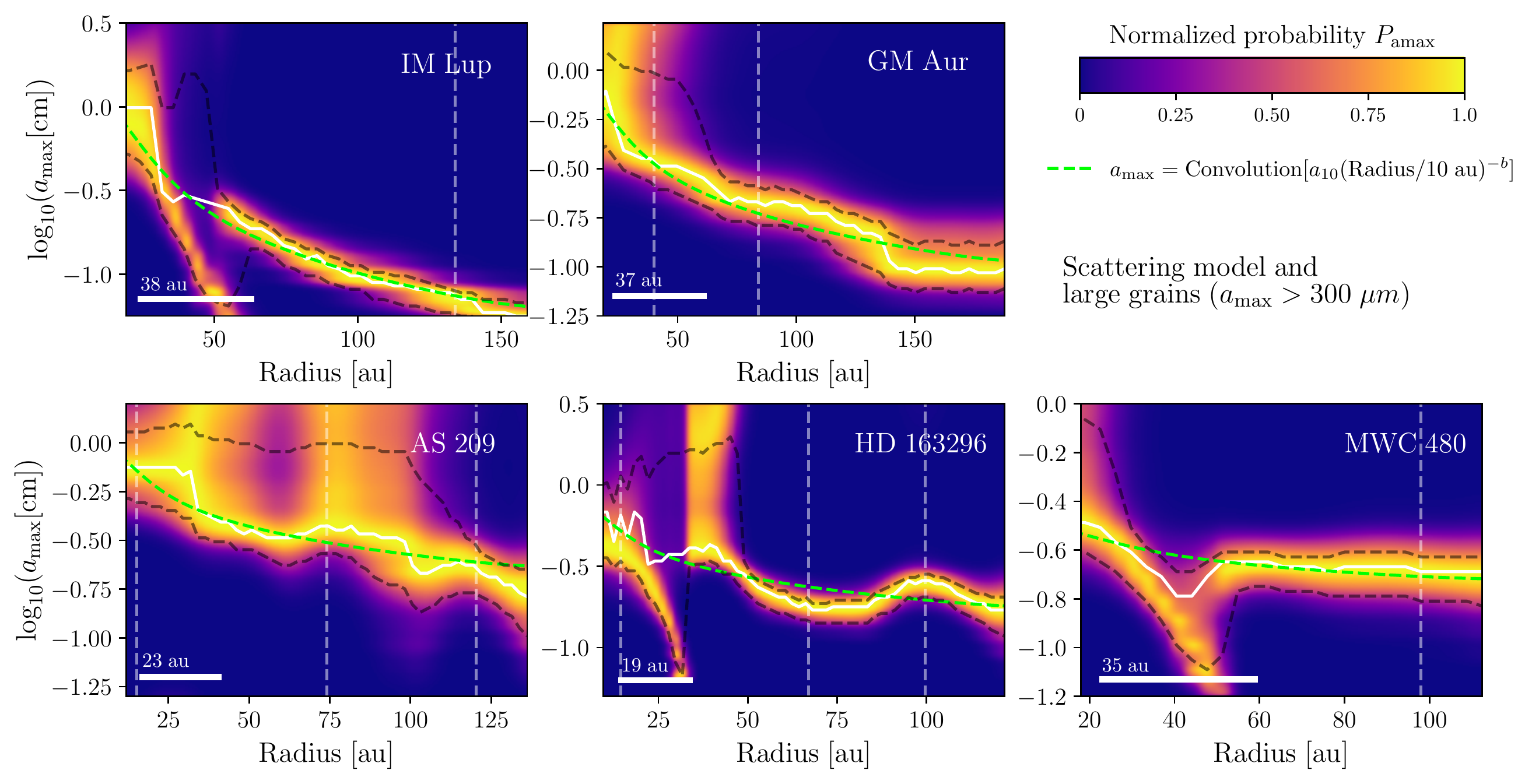}
    \caption{Probability distribution of the maximum grain size for all disks in our sample when scattering is taken into account in the radiative transfer model and the large-grain solution ($a_{\rm max} > 300 \ \mu$m) is considered for the inner disk. The white solid line is the best-fit maximum grain size ($a_{\rm best}$). The color scale is the marginal probability distribution $P_{a_{\rm max}}$. Dashed black lines are the 1$\sigma$ uncertainties,  and the horizontal white line in the lower left corner is the beam size of each disk. The green dashed line is a power-law fit to the maximum grain size with radius; vertical dashed lines mark the position of the bright rings.}
    \label{fig:Results_amax_Scat}
\end{figure*}

Figures \ref{fig:Results_sigma_Scat} and \ref{fig:Results_amax_Scat} show the dust surface density and maximum grain size, respectively, from the scattering case and large grains. The white solid lines are the best values ($\Sigma_{\rm best}, a_{\rm best}$), and the color scales are the marginal probabilities ($P_{\Sigma}$, $P_{\rm amax}$).
The dust masses and the free parameter in the power-law fit to $a_{\rm best}$ are also summarized in Table \ref{tab:Dust_inf}.

\begin{figure*}[ht!]
    \centering
    \includegraphics[width=\textwidth]{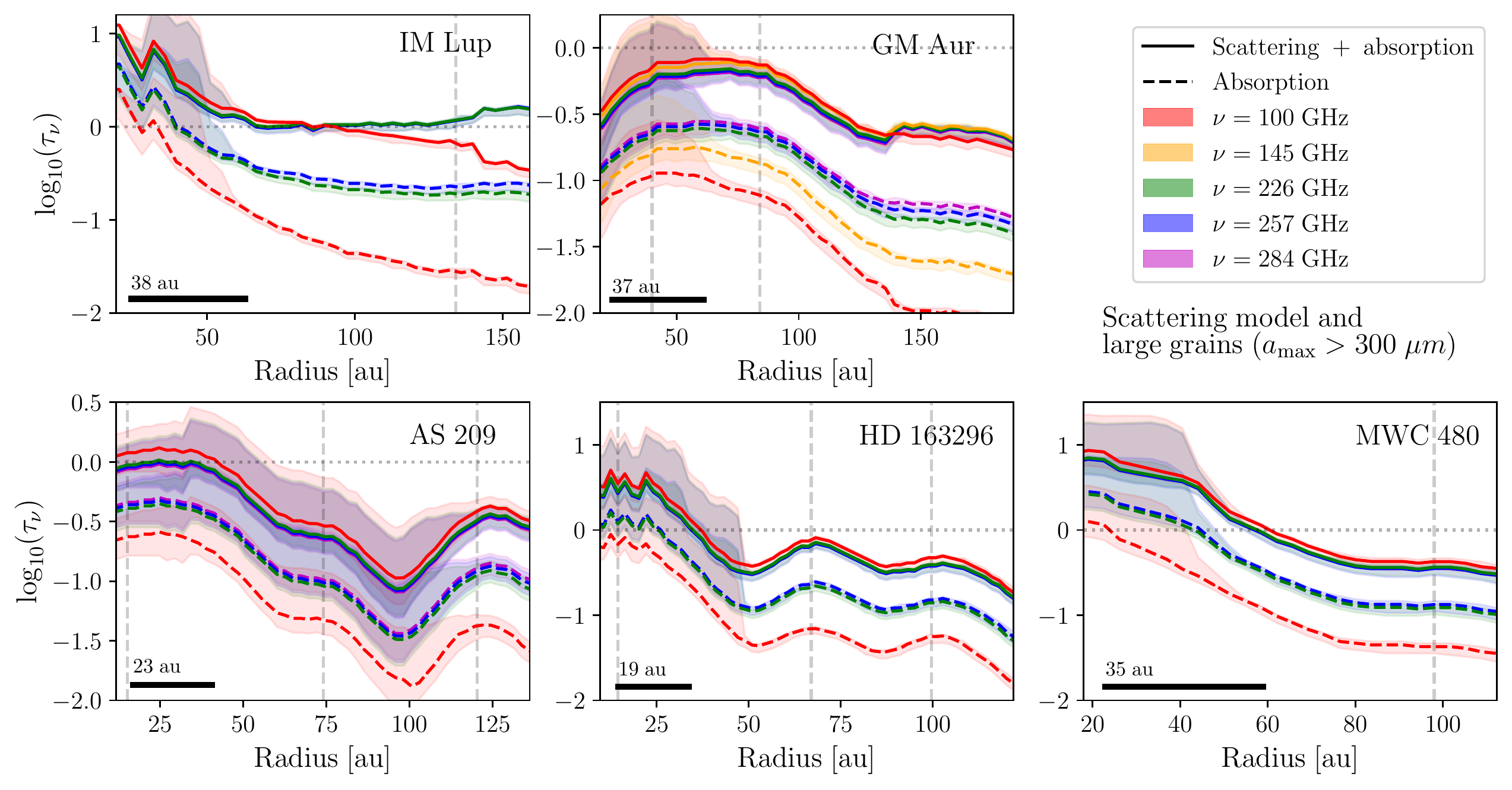}
    \caption{Dust optical depths at 100 GHz (red), 145 GHz (orange), 226 GHz (green), 257 GHz (blue), and 284 GHz (magenta) of the five disks (see legend in the upper right corner of each panel) in the scattering models and large grains ($a_{\rm max} > 300 \ \mu$m). The colored dashed and solid lines are the optical depths associated with absorption and scattering + absorption, respectively. The shaded areas are the uncertainties of each profile. 
    The horizontal black line in the bottom-left corner is the beam size of each disk. Vertical dashed lines mark the position of the bright rings, and the horizontal dotted line is a reference value where the optical depth is 1.}
    \label{fig:optical_depth_Scat}
\end{figure*}

Figure \ref{fig:optical_depth_Scat} shows the optical depths associated with the absorption coefficient ($\tau_{\nu}^{\rm abs} = \Sigma_{\rm d} \kappa_{\nu}$), and the total optical depth, where scattering opacity is also included ($\tau_{\nu}^{\rm sca} = \Sigma_{\rm d} \kappa_{\nu}/(1-\omega_{\nu})$), for the five disks at different wavelengths. As the albedo is large, total opacities increase by one order of magnitude compared to the nonscattering opacities. The Toomre $Q$ parameter and the maximum Stokes number for the scattering case are quite similar to those derived when scattering is neglected (Figure \ref{fig:toomre_stokes}), since the maximum grain size and dust density have similar behaviors and magnitudes; thus, they are not presented again.


On the other hand, the solution for small grains ($a_{\rm max} < 300 \ \mu m$) is particularly interesting because if the maximum grain size is a few hundred micrometers, the disks could be consistent with the polarization pattern observed in many disks \citep{Kataoka_2015}.

\begin{figure*}[ht!]
    \centering
    \includegraphics[width=\textwidth]{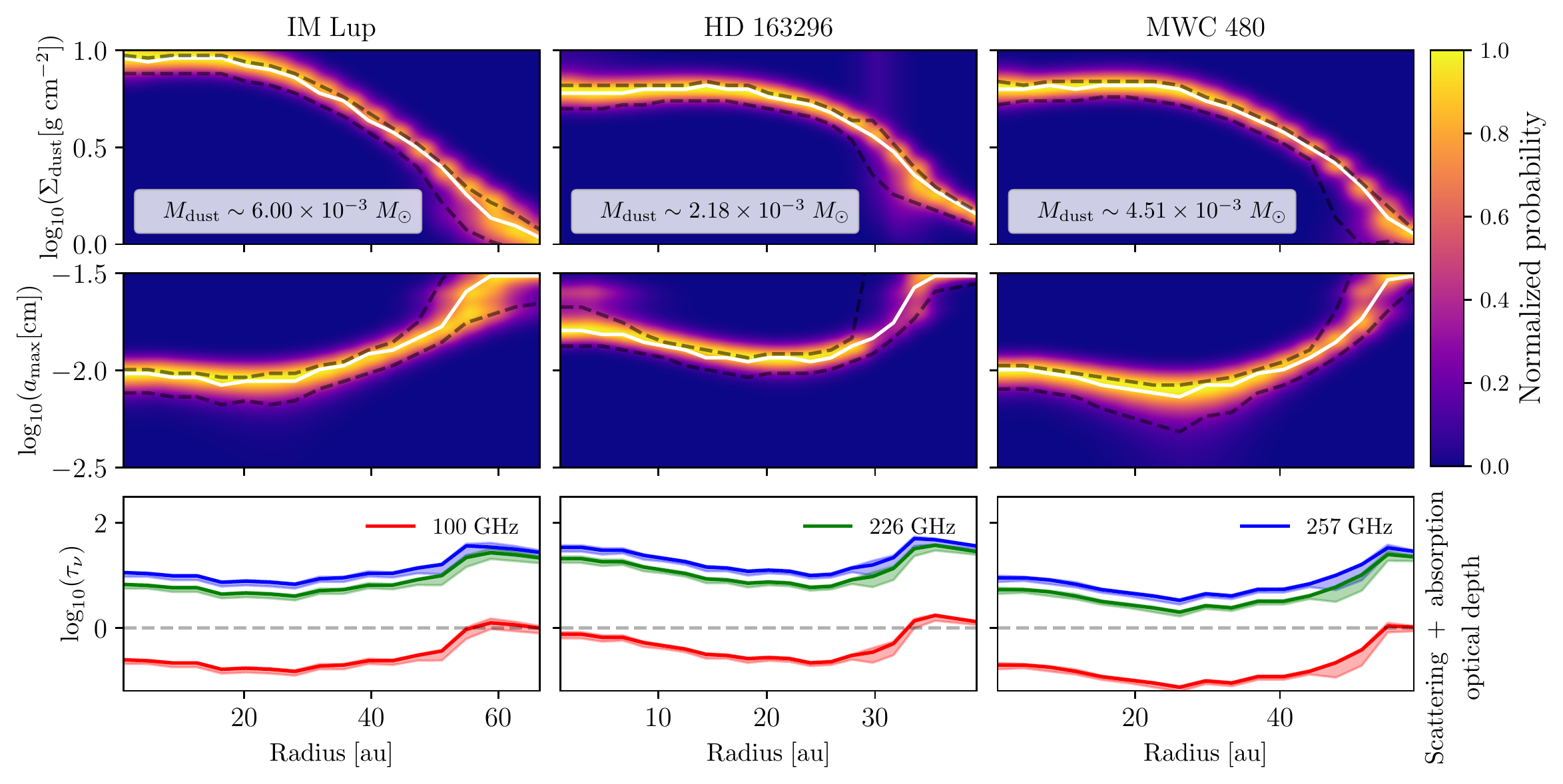}
    \caption{Probability distribution of the dust surface density (top panels), maximum grain size (middle panels), and optical depths (bottom panels) for three disks in our sample when scattering is taken into account in the radiative transfer model and the small-grain solution ($a_{\rm max} < 300 \ \mu$m) is considered for the inner disk. 
    The color scales in the top and middle panels are the marginal probabilities for each parameter, and the black dashed lines are the 1-sigma uncertainties. The dust mass (only within the radial extent where a solution is found) of each disk is shown in the legend of the top panels.
    The optical depths in the bottom panels consider scattering and absorption. The horizontal dashed line in the bottom panels is a reference value where the optical depth is 1. The shaded areas are the uncertainties of each profile. Different colors correspond to different frequencies (see colored legends).}
    \label{fig:small_grain_solution}
\end{figure*}

Figure \ref{fig:small_grain_solution} shows the small-grain solution for three of the disks, as GM Aur and AS 209 have absorption optical depth smaller than 1 at all radii (Figure \ref{fig:optical_depth_Scat}) and thus are not consistent with this small-grain solution. We emphasize that scattering has important effects on the emergent intensity only when the optical depth associated with the absorption component is thick \citep[][]{Zhu_2019, Sierra_2020}.
The top panels of Figure \ref{fig:small_grain_solution} are the dust surface densities, the middle panels are the maximum grain sizes, and the bottoms panels are the optical depths considering scattering and absorption opacity. The radial extent of each disk is shown according to the region where a solution with some-hundred-micrometer grain sizes is found.
This small-grain solution requires a large amount of solids. This occurs because the dust opacity of $\sim 100\ \mu$m grains at millimeter wavelengths is smaller than the dust opacity of millimeter grains. For example, at $\lambda = 1.3$ mm, the dust opacity of 1 mm grains is a factor of $\sim 4.5$ larger than the opacity of $100 \ \mu$m grains.
Then, the small grains need to have a larger amount of mass in order to reach the same intensity of millimeter grain sizes. The dust mass (only within the radius shown in Figure \ref{fig:small_grain_solution}) is shown in the lower left corner of each panel, which is a factor of 2-5 larger than the whole mass associated to millimeter grains (Table \ref{tab:dust_continuum_prop}).

\begin{figure*}[ht!]
    \centering
    \includegraphics[width=\textwidth]{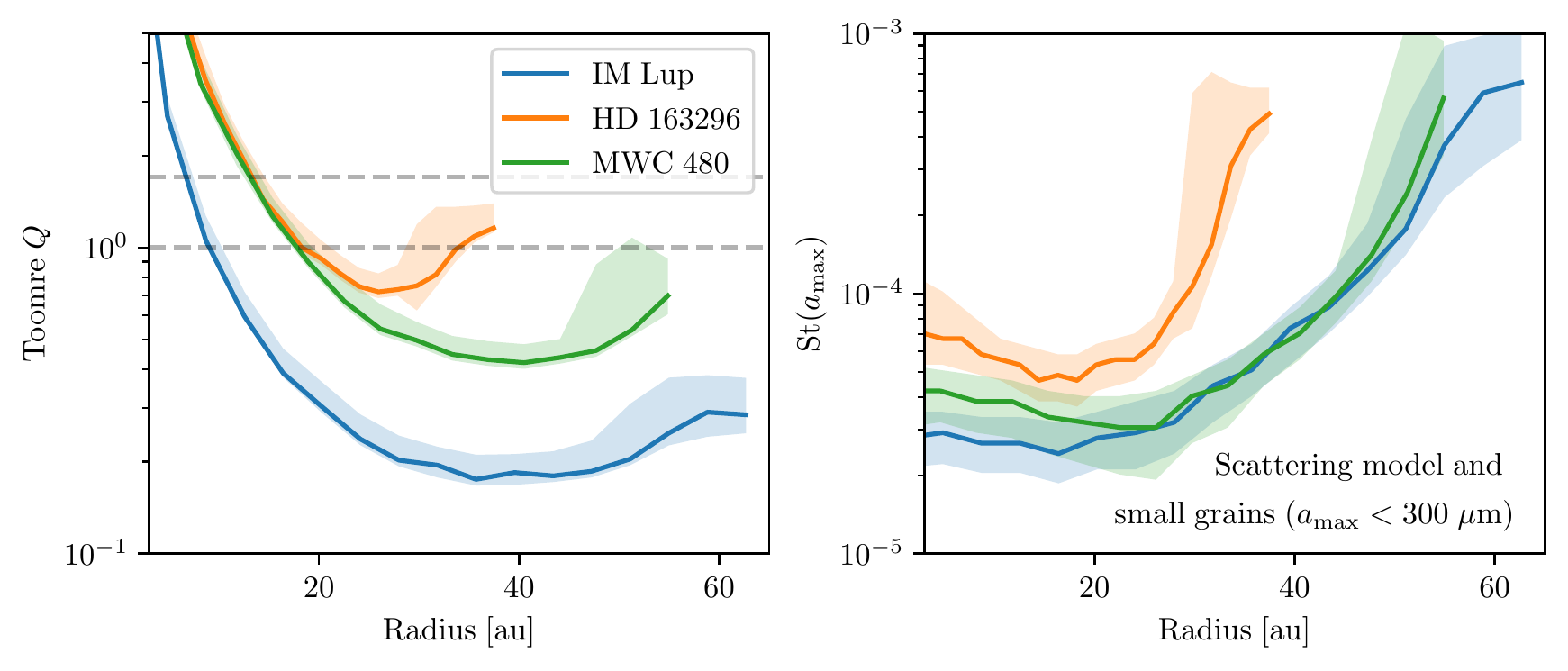}
    \caption{Toomre $Q$ parameter (left) and maximum Stokes number (right) for three of the disks (see legend on the left panel) in the scattering model with small grains ($a_{\rm max} < 300 \ \mu$m). The horizontal dashed lines in the left panels ($Q <$ 1,  1.7) delimit the regions where the disk becomes gravitationally unstable to linear and second-order perturbations, respectively. The shaded areas are the uncertainties of each curve.}
    \label{fig:toomre_stokes_small}
\end{figure*}

Figure \ref{fig:toomre_stokes_small} shows the Toomre parameter (left panel) and the maximum Stokes number (right panel) for the small-grain solution and the three disks. The Toomre parameter and the maximum Stokes number are small compared with those in Figure \ref{fig:toomre_stokes}, since the maximum grain sizes are smaller and thus the dust surface densities are higher compared with the results from the large-grain solution.

\section{Discussion} \label{sec:discussion}
Radial profiles of the dust surface density, maximum grain size, optical depths, Toomre $Q$ parameter, and maximum Stokes number of the five disks were computed in Section \ref{sec:results}. Here we discuss the results from the nonscattering (Section \ref{subsec:no-scattering}) and scattering (Section \ref{subsec:scattering}) models, discuss the main properties of each disk, and compare our results with constraints from previous works. 
However, before discussing our results, it is important to address how they depend on the assumptions and modeling.

First, all the inferred properties depend on angular resolution, which in our sample varies from 19 to 39 au. The contrasts between rings and gaps are expected to be higher than those presented in this work if the rings and gaps are not resolved at all wavelengths. Beam averaging from optically thick regions (with low spectral index) and optically thin regions (with large spectral index) could lead to overestimation of the spectral indices at the location of rings and simultaneously lead to underestimation of $\alpha$ in gaps \citep[e.g. see an example and discussion in ][]{Lin_2020}. Multiwavelength observations that could spatially resolve the dust rings at all wavelengths could better resolve these different regions and better infer their dust properties.

The dust surface density and maximum grain size are the main properties from which all other parameters are computed. The maximum grain size is assumed to be constant above the midplane, i.e. we do not consider settling. The magnitude of the maximum grain size depends on the assumed slope of the particle size distribution ($p$). In particular, we choose $p=2.5$, which is a typical value for the slope when the maximum grain size is limited by radial drift \citep{Birnstiel_2012}, and gives a lower limit to the inferred maximum grain size, as we show in Section \ref{sec:methodolody}. Recently, \cite{Macias_2021} fitted the dust properties of the disk around TW Hya. In their modeling, the value of $p$ was initially fixed and then it was considered as a free parameter. In the former case, the maximum grain size distribution has local maxima and minima that correlate with the dust rings and gaps. In the latter case, most of the radial substructure of the maximum grain size disappears, but the radial profile for $p$ exhibits local minima and maxima at rings and gaps, respectively. 
The maximum grain size and the slope $p$ are highly degenerate. A more realistic model (e.g. dynamical dust simulations) should be able to account for changes in $p$ in gaps and rings. However, this is beyond the scope of this paper.

On the other hand, as discussed in Section \ref{sec:methodolody}, the dust temperature model was fixed for each disk using the midplane temperature from Zhang et al. (2020). This avoids degeneracy, especially in the optically thin regime, where the emergent intensity scales as $I_{\nu} \propto \Sigma_{\rm d} T_{\rm d}$, and one can only constrain the product between the temperature and dust surface density. However, uncertainties in the emergent intensities are propagated to the dust surface density and maximum grain size, and in some cases the observed multiwavelength emission can be reproduced using a large range of dust surface densities and maximum grain sizes, even if these two are the only free parameters. More details about this degeneracy are described in Appendix \ref{app:degeneracy}.

The optical depths are directly computed from the dust surface density and from the maximum grain size, which sets the opacity coefficient and albedo in the scattering case. Thus, there are no additional assumptions when computing the optical depth of the dust continuum emission.

However, when computing the Toomre $Q$ parameter and the maximum Stokes number, additional assumptions about the gas properties are needed. First, the gas surface density is assumed to be 100 times the dust surface density. This factor comes from the typical value in the ISM. However, \cite{Ansdell_2016} computed the dust mass from 890 $\mu$m continuum observations and the gas mass from CO isotopologue in a survey of disks in Lupus and found that most disks in this region have a dust-to-gas mass ratio ($\epsilon$) larger than 1/100, even when specific examples with smaller dust-to-gas mass ratio have also been inferred \citep[e.g. in the TW Hydra disk;][]{Zhang_2017}.

Additionally, the disk sizes obtained from the millimeter dust continuum emission \citep[see a complete analysis of the MAPS disks sizes in][]{Law_2020_rad} tend to be smaller than that obtained from the gas molecular line emission \citep[e.g.,][]{ Pietu_2005, Isella_2007}. Then, it is expected that most of the continuum emission at these wavelengths comes from regions where the dust-to-gas mass ratio has been enhanced with large grains by radial drift. 

Numerical simulations of dust trapping in rings \citep[e.g.,][]{Gonzalez_2017, Pinilla_2020}, or in spiral arms \citep[e.g.,][]{DiPierro_2015b} have also found changes in the dust-to-gas mass ratio as dust grains migrate toward pressure maxima. However, the magnitude of $\epsilon$ depends on factors such as the amount of solids in the simulation, turbulence, fragmentation velocity, the evolutionary stage (all poorly constrained), or whether the dust back reaction is taken into account or not. In these works, $\epsilon$ varies from $\sim 0.003$ to $\sim 1$.

As the Toomre parameter and Stokes number are inversely proportional to the gas surface density ($Q, \rm{St} \propto \Sigma_{\rm g}^{-1} = \epsilon \Sigma_{\rm d}^{-1}$), they are only vertically shifted with respect to the results in Figure \ref{fig:toomre_stokes} (where we assume $\epsilon = 1/100$) if the dust-to-gas mass ratio is modified. For example, in \cite{Dullemond_2018}, they assume that the upper limit of the gas surface density is the dust surface density (i.e. $\epsilon = 1$). In that upper limit, the Stokes numbers in Figure \ref{fig:toomre_stokes} lie within $1$ and $10$, and all the disks would be gravitationally stable.
We also compute the Stokes number using the gas surface density derived from the CO column densities by \cite{Zhang_2020}, and we find that the Stokes numbers for the five disks lie between $10^{-2}$ and $10^{-1}$, as in Figure \ref{fig:toomre_stokes}.

Thus, scaling the dust mass by a factor of 100 is only a first approximation to the gas mass. Many physical processes can change this value and modify the Toomre parameter and the maximum Stokes number in Figure \ref{fig:toomre_stokes}.
In addition, the factor of 100 is assumed constant thought the disk, i.e. the shape of the gas and dust surface density are the same; however, this approximation is valid for disks with grains well coupled with the gas and/or a high turbulence \citep[e.g.,][]{Birnstiel_2013, Ruge_2016}, such that these solids can be used to trace both the gas and dust surface density. 
Smoother structures are expected in gas than in dust. On average, the widths of all gas features for the MAPS disks are much larger than the dust, and the chemical substructures have lower relative contrasts than the continuum substructures \citep{Law_2020_rad}.
Then, one expects that the dust-to-gas mass ratio should be larger in rings (where dust grains are being trapped) than in gaps (where dust grains are depleted). Consequently, the assumption $\Sigma_{\rm g} = \Sigma_{\rm d}/\epsilon$ with a constant $\epsilon$, overestimates the gas surface density in rings and underestimates it in gaps. Thus, the Toomre parameter and maximum Stokes number shown in Figure \ref{fig:toomre_stokes} are overestimated in gaps and underestimated in rings. 

On the other hand, the sound speed is computed from the dust temperature in the midplane, which is assumed to be the same as the gas temperature. However, the two components may not be in thermal equilibrium. Finally, we assume that the gas velocity is Keplerian, but the gas velocity is sub-Keplerian or super-Keplerian depending on the local pressure gradients \citep{Takeuchi_2002}. Even when a deviation from Keplerian angular velocity is smaller than 10\% \citep[e.g.][]{Teague_2018, Teague_2019}, it produces a sufficiently large observational signature. In fact, \cite{Rosotti_2020} were able to estimate the width of the gas rings from the velocity curves in a disk with deviations of only $\sim 3 \%$ in Keplerian velocity. Additional uncertainty is introduced owing to a lack of precise knowledge in the stellar masses or the assumption of equal dust composition and structure throughout the disks.

The dust masses estimated in this work are directly computed from integrating the dust surface densities. As mentioned in Section \ref{sec:methodolody}, the dust surface density is computed inside the radius where the enclosed flux at Band 3 is 95\% of the total flux. However, the continuum emission at higher frequencies is more extended than this radius, but the mass beyond this radius is not considered in the total mass estimate. Table \ref{tab:FluxPercentage} shows the percentage of the total flux at different frequencies within the radius that encompasses 95\% of the total flux in Band 3. At higher frequencies, the enclosed flux for GM Aur and AS 209 is larger than 90\%, while for HD 163296 and MWC 480 it is larger than 85\%. For IM Lup, the most massive disk, the enclosed flux fraction is only 78\%. Even when the dust masses inferred in this work could be interpreted as a lower limit to the total dust mass, the inferred dust surface densities decreases by more than one order of magnitude from the inner to the outer disk, and thus a nonsignificant contribution to the total the mass is expected beyond the fitting radius.

\begin{table*}[ht!]
    \centering
    \caption{Percentage of the Enclosed Flux within the Fitting Radius, Defined as the Radius Where the Enclosed Flux at Band 3 Is 95\% of the total Flux.} 
    \begin{tabular}{c|ccccc}
        \hline \hline
        Frequency & \multicolumn{5}{c}{Percentage} \\
         (GHz) & IM Lup & GM Aur & AS 209 & HD 163296 & MWC 480 \\
         \hline 
        100  & 95\%  & 95\%   & 95\%   & 95\%      & 95\%  \\
        145  & ...   & 93\%   & ...    & ...       & ...   \\        
        226  & 78\%  & 91\%   & 91\%   & 84\%      & 88\%  \\
        257  & 78\%  & 91\%   & 93\%   & 85\%      & 88\%  \\
        284  & ...   & 91\%   & 93\%   & ...       & ...   \\
        \hline 
        Fitting radius & 160 au & 188 au & 136 au & 122 au & 113 au\\
        \hline
    \end{tabular}\\
    Note: These percentages were computed using the radial profiles from the CLEANed images at the same angular resolution (see Figure \ref{fig:intensity_profiles}).
    \label{tab:FluxPercentage}
\end{table*}

In general, the dust masses agree or are within a factor of a few compared with previous estimates (each disk is discussed below). The reasons for the disagreement (in addition to the fitting radius) are diverse, but the main are the following: Typically, the dust mass is estimated from the millimeter flux as
\begin{equation}
    M_{\rm dust} = \frac{F_{\nu} d^2}{\kappa_{\nu} B_{\nu}(T_d)},
    \label{eq:DustMass}
\end{equation}
where $F_{\nu}$ is the flux at some frequency $\nu$, $d$ is the distance to the source, $\kappa_{\nu}$ is the opacity coefficient at the same frequency, and $B_{\nu}$ is the Planck function at temperature $T_d$. Equation \ref{eq:DustMass} assumes that the disk is optically thin everywhere, which is not always valid in dense rings or the inner disk, especially at short millimeter wavelengths (e.g., at ALMA Band 7 or Band 6). The opacity coefficient depends on the dust composition assumed, but in general the dust opacities from \cite{Beckwith_1990} or \cite{Dalessio_2001} have been most used in the literature. And recently, the DSHARP opacities \citep{Birnstiel_2018} have become the standard to compute opacity properties. In particular, the \cite{Beckwith_1990} opacity prescription was found to be consistent with the dust opacities inferred at ALMA Bands 7, 6, and 3 in the disk around HH 212 \citep{Lin_2021}.
The absorption spectrum from \cite{Beckwith_1990} is given by $\kappa_{\nu} = 2.3 (\nu/ 230 \ {\rm GHz}) ^{\beta} {\rm cm}^2 \ {\rm g}^{-1}$, and a value of $\beta = 1$ is usually assumed to scale this opacity to other frequencies. For example, the opacity coefficient from Beckwith at 100 GHz is 1.0 ${\rm cm}^2 \ {\rm g}^{-1}$. From the DSHARP opacities, the opacity coefficients for a grain size of 1 mm at 230 and 100 GHz are 1.89 and $0.31 \ {\rm cm}^2 \ {\rm g}^{-1}$, respectively. At 230 GHz, the opacity coefficient only differs by a factor of 1.2, but at 100 GHz, they differ by a factor of 3.2. The main reason for the disagreement is the assumed value of $\beta$, which depends on the grain size and can only be estimated given the spectral index between two wavelengths. 
In our modeling, the disks are not assumed to be optically thin, and the value of $\beta$ is adopted depending on the maximum grain size at each radius.

Thus, differences in dust composition, temperature, and nonresolved optically thick emission can be some of the reasons of disagreement between the inferred dust masses in our works and others.

\subsection{Scattering versus Nonscattering} \label{subsec:ScatvsNon}
Scattering is expected to modify the spectral indices for optically thick emission (\citealt{Liu_2019}, \citealt{Zhu_2019}, \citealt{Sierra_2020}). In particular, spectral indices below 2 can be obtained in the optically thick regime if the albedo is high and increases with frequency. The latter condition is satisfied in ALMA Bands 7, 6, and 3 for spherical dust grains with sizes of about a hundred micrometer.
For grains larger than 1 mm, scattering cannot reproduce spectral indices below $\sim 2.1$ (see Figure 9 in \citealt{Zhu_2019} and Figure 3 in \citealt{Sierra_2020}), while in the nonscattering case the optically thick spectral index is always 2.

These previous effects were computed in the Rayleigh-Jeans (RJ) regime. However, this regime is not valid for low dust temperatures. The RJ regime is only valid for $[\nu /{\rm GHz}] << 20 [T/ {\rm K}]$, which is not fully satisfied at Band 6 given the dust temperature models in our disk sample (Appendix \ref{app:Temperature}). When the RJ regime is not satisfied, the spectral indices (in both the scattering and nonscattering cases) decrease because the peak of the Planck function is displaced to the submillimeter range according to Wien's law \citep[e.g. see Figure A1 in][]{Sierra_2020}.

Whether or not RJ is valid, the change of the optically thick spectral index with maximum grain size in the scattering case is responsible for the two possible solutions found in the inner disks. In particular, the optically thick spectral indices for few-hundred-micrometer and for millimeter grains sizes are similar.

The solutions for the nonscattering case and the scattering case with large grains ($a_{\rm max} > 300 \ \mu$m) are the same in the outer disk. This occurs because the grains are optically thin with respect to absorption at these radii, and the radiative transfer solution with scattering (Equation \ref{eq:Intensity_scatI}) reduces to the nonscattering solution (Equation \ref{eq:Intensity_no_scat}) in this limit, even if the albedo is high.
By contrast, the solutions in the inner disks do depend on whether scattering is included or not. The best models are dominated by millimeter grain sizes in the scattering and nonscattering case, and the dust surface densities are also similar. However, the marginal distributions are modified, due to changes of the spectral index with grain size in this regime. From Table \ref{tab:dust_continuum_prop}, one can see that the dust masses are similar in the scattering and nonscattering models, except for IM Lup, where the dust mass in the scattering model is a factor of 2 larger than the dust mass in the nonscattering model. This occurs because if scattering is neglected, the dust mass from the nonscattering model is underestimated in the optically thick regime \citep{Zhu_2019}, and IM Lup is the most optically thick disk in our sample in Band 6.

The small-grain regime ($a_{\rm max} < 300 \ \mu$m) in the scattering case provides an alternative solution that is able to reproduce the observations in the inner disks. No solution is found in the outer disks in this case because the individual regions that can explain each wavelength do not overlap when $a_{\rm max} < 300 \ \mu$m. The solution in the inner disks corresponds to grain sizes between 100 and 300 $\mu$m, and the dust surface densities are higher than the large-grain solution. This occurs because the opacity coefficient of hundred-micrometer grains is smaller than the opacity of millimeter grains by a factor of $\sim 5$, i.e. more mass is required for small grains that have a smaller opacity coefficient, such that they can have the same amount of emission of larger grains with a larger opacity coefficient. 

High dust surface densities and some hundred-micrometer sizes were already suggested by \cite{Ueda_2020} and \cite{Macias_2021} in the inner region of the TW Hya disk. They fitted the SED using a scattering and a nonscattering model, and found that the disk is consistent with 300 $\mu$m grains and large densities ($\Sigma_{\rm d} = 10$ g cm$^{-2}$) if scattering is taken into account. However, if they neglect the scattering effects, the SED could also be explained by optically thick emission and large grains.

Some of the disks in our sample have been studied with polarized observations. For example, \cite{Hull_2018} found that polarized properties of the inner disk of IM Lup are consistent with 61 $\mu$m grain sizes, but millimeter to centimeter grain sizes are needed to reproduce the unpolarized spectrum (consistent with this work).
The polarization fraction of the inner disk (0.\arcsec 5) of AS 209 at 870 $\mu$m is only $~0.2$\% \citep{Mori_2019}, a very low value compared with the expected polarized fraction if the maximum grain size is around some hundred micrometers \citep{Kataoka_2015}. This supports the fact that we find that some-hundred-micrometer sizes are not consistent with the SED in the inner disk, and millimeter grains are needed.
\cite{Ohashi_2019} used the polarized observations of HD 163296 \citep{Dent_2019} to model the dust content within the disk. They found that the gaps should have a large contribution of some-hundred-micrometer sizes to explain the polarimetric observations; however, the rings could have no contribution of these grains but much larger grains.
No polarized emission at low angular resolution (2$.\arcsec$0-2$.\arcsec$6) was detected at Band 7 for GM Aur and MWC 480 \citep{Hughes_2013}; however, \cite{Harrison_2019} detected polarized emission parallel to the disk minor axis in the inner disk of MWC 480, with a polarization fraction of $\sim 1 \%$, which can be explained by self-scattering of large grains ($>1$ mm). More multiwavelength polarized observations are needed to model and discern the maximum grain sizes in these disks.

In particular, the small-grain solution has two important problems. First, it requires a large dust mass to produce sufficient millimeter emission, and in that case, the optically thick inner disks of IM Lup, HD 163296, and MWC 480 would be gravitationally unstable if the gas-to-dust ratio is 100, as shown in the left panel of Figure \ref{fig:toomre_stokes_small}. The maximum Stokes numbers of these grains (right panel of Figure \ref{fig:toomre_stokes_small}) are $\sim 2$ orders of magnitude smaller compared with the large-grain solution, and they would be perfectly coupled with the gas, making it difficult for the sharp ring structures observed in continuum emission to exist. Second, this regime is only able to explain the inner disk. Only millimeter-sized particles can explain the outer disk emission, providing continuous dust profiles from the inner to the outer disk, with no sharp transition.

Despite these physical problems associated with the small-grain solution, it is consistent with the spherical grain sizes expected from the polarization patterns observed in many disks \citep[e.g., ][]{Kataoka_2015, Bacciotti_2018, Hull_2018, Mori_2019}, which are mainly tracing the unresolved emission from the inner disks. However, since the polarization degree of oblate dust grains is higher than that from spheres \citep{Kirchschlager_2020}, the large-grain solution may also be consistent with the observed polarization properties. 

\begin{figure*}[ht!]
    \centering
    \includegraphics[width=0.95\textwidth]{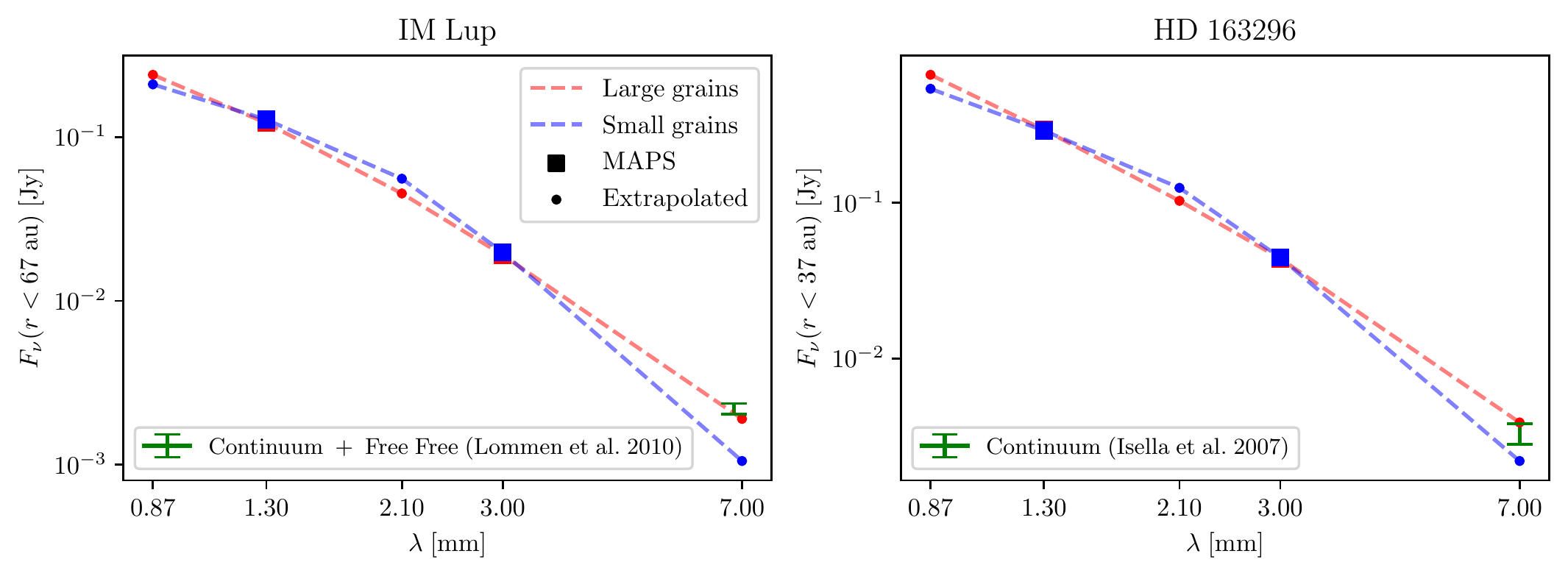}
    \caption{SED for IM Lup (left) and HD 163296 (right). The fluxes are measured within the radius where the small-grain solution is found. The squares are the MAPS observations, while the circles are the predicted fluxes from the large (red) and small (blue) grain models. The green error bar is the nonresolved flux measured at long wavelength.}
    \label{fig:SED_7mm}
\end{figure*}

Observations at larger wavelengths can also be used to discern between the small- and large-grain solutions, even if the observations are nonresolved. The expected 7 mm flux from the large- and small-grain models differs by a factor of two. For example, the estimated IM Lup flux at 7 mm for the large- and small-grain models is $\sim 1.9$ and 1.1 mJy, respectively, and for HD 163296, the fluxes are $\sim 3.9$ and 2.2 mJy, respectively.
Using the Australian Telescope Compact Array (ATCA), \cite{Lommen_2010} reported a total flux of $2.2 \pm 0.16$ mJy at 6.8 mm for IM Lup, and using the Very Large Array (VLA), \cite{Isella_2007} reported a total flux of $4.5\pm 0.5$ mJy for HD 163296. In the latter work, they also determined that the free-free contribution at this wavelength is $1.2$ mJy ($\sim 27\%$), such that the dust continuum flux of HD 163296 at 7 mm is $3.3 \pm 0.5$ mJy.
Figure \ref{fig:SED_7mm} shows the SED for IM Lup and HD 163296 for the large- and small-grain solution and their observed flux at 7 mm. For IM Lup, the free free contribution at 7 mm has not been estimated, but it should be $\sim 14 \%$ or $50 \%$ of the total flux in order to be consistent with the large- or small-grain model, respectively.
For HD 163296, the 7 mm flux from the large-grain solution is slightly larger than the observed continuum flux and differs from the small-grain model by a factor of 1.5. Thus, the total flux at 7 mm provides an additional constraint that suggests that the maximum grain sizes around HD 163296 are millimeter sized.

Future polarized observations at long wavelengths (e.g. $\lambda=$ 7 mm) will also help us to constrain grain sizes in the inner disks, and more complex modeling should be formulated to explain the whole structure of the inner disk with scattering, for example, a two-layer model with small grains in the disk surface (to explain the spectral indices) and large grains in the midplane (dominating the dust opacity and mass). 

Finally, note that the optical depths with scattering are larger than the nonscattering case by one order of magnitude or more (Figure \ref{fig:optical_depth_Scat}). This large increase occurs because the albedo of millimeter grains at radio frequencies is large \citep[$\omega_{\nu} > 0.9$; ][]{Birnstiel_2018}, and the total optical depth increases by a factor of $1/(1-\omega_{\nu})> 10$. The disks tend to be close to the optically thick regime ($\tau_{\nu} = 1$) when scattering opacity is turned on.
In the small-grain solution (bottom panels in Figure \ref{fig:small_grain_solution}), the optical depths at Band 6 are optically thick, while the optical depth at Band 3 is optically thin in all disks.
In the next sections we analyze the results from each disk, focusing on the dust properties derived from the nonscattering model, and the scattering model with large grains.

\subsection{Source specific}
\subsubsection{IM Lup}\label{subsec:IMLup}
IM Lup is a $\sim$ 1.0 Myr old T Tauri star with a mass of 1.1 $M_{\sun}$ \citep{Andrews_2018}. 
It is the only disk in our sample with spiral arms traced in the millimeter dust continuum. A complete study of its morphology was done by \citet{Huang_2018b} using high angular resolution observations with ALMA at Band 6 (44 mas or $\sim4$ au resolution). They also found a ring at 134 au, which is plotted as a vertical dashed line in our figures as a reference. In addition, \cite{Law_2020_rad} found a dark and a bright ring at 209 and 220 au, respectively.

This disk has also been studied by multiwavelength observations (optical, near-infrared, and millimeter emission) by \cite{Pinte_2008}. They determined that dust grains are millimeter sized in the disk midplane (consistent with our results) and suggest that the disk may be gravitationally unstable.

We found that the dust surface density (Figure \ref{fig:Results_sigma}) monotonically decreases from $\sim 3$ g cm$^{-2}$ in the inner disk ($20 \sim $au) to $\sim 0.1$ g cm$^{-2}$ in the outer disk ($\sim 140$ au), while the maximum grain size (Figure \ref{fig:Results_amax}) decreases from $\sim 1$ cm in the inner disk to several hundred micrometers in the outer disk. The dust mass is $1.92 ^{+1.91}_{-0.46} \times 10^{-3} M_{\sun}$ in the nonscattering model, and $3.64 ^{+1.99}_{-1.41} \times 10^{-3} M_{\sun}$ in the scattering model with large grains. The former is close to the dust mass of $0.7 \times 10^{-3} M_{\sun}$ in \cite{Pinte_2008} and similar to $1.6 \times 10^{-3} M_{\sun}$ in \cite{Cleeves_2016} \footnote{The dust masses reported in previous works are updated using the distances summarized in Table 1 in \cite{Oberg_2020}.}.

The optical depth of the emission in IM Lup appears optically thick within $\lesssim 30$ au and optically thin beyond this radius (Figure \ref{fig:optical_depth}). When scattering is taken into account (Figure \ref{fig:optical_depth_Scat}), the total opacity is close to or above 1 for all wavelengths at all radii.
The Toomre $Q$ parameter (Figure \ref{fig:toomre_stokes}) is close to or below 1.7 at all radii, indicative of a gravitationally unstable system. Such a small value is expected given the observed spiral arm structure in Band 6 (Figure \ref{fig:continuum_MAPS}).
This is the only disk in our sample where the Stokes number decreases with radius beyond $\sim$ 75 au, and its overall value is small ($1 \times 10^{-2}$ and $3 \times 10^{-2}$). This behavior occurs because of the high surface density, which tends to a constant value beyond $\sim$ 100 au, while the maximum grain size monotonically decreases with radius. Such small values of the Stokes number are expected in dust trapping models by spiral arms when the fragmentation velocity is set to 10 m s$^{-1}$  \citep{DiPierro_2015b}.

\subsubsection{GM Aur} \label{subsec:GMAur}
GM Aur is a T Tauri star with an age within $\sim$ 3 and 10 Myr and a mass of 1.1 $M_{\sun}$ \citep{Macias_2018}. 
The GM Aur disk is a transition disk with a large inner cavity previously reported by many authors \citep[e.g.,][]{Hughes_2009, Andrews_2011, Guilloteau_2011, Hornbeck_2016} and recently studied using multiwavelength observations by \cite{Macias_2018} and \cite{Huang_2020} at high angular resolution ($\sim 8$ au in ALMA and $\sim 26$ au in VLA). They found two rings at 40 and 84 au, which are plotted as vertical dashed lines in our figures as a reference.

The spectral indices of this disk have local minima at the ring position (Figure \ref{fig:spectral_indices}). However, because of the low angular resolution in Band 3 and beam averaging of Band 6, it is not possible to clearly distinguish high contrasts between rings and gaps. When the spectral index is computed between Bands 4 and 6 using higher angular resolution profiles, the contrast between rings and gaps is clearer, as shown in \cite{Huang_2020}.
The dust surface density (Figure \ref{fig:Results_sigma}) increases within the first 40 au, reaching a maximum value in the inner ring position, and then it decreases toward the outer disk, with a local maximum in the outer ring. The maximum grain size (Figure \ref{fig:Results_amax}) monotonically decreases from the inner disk, where $a_{\rm max} \sim 8$ mm, to the outer disk, where $a_{\rm max} \sim 1$ mm. 
GM Aur is be optically thin for the entire disk (Figure \ref{fig:optical_depth}). The largest optical depth is reached in the inner ring at Band 7, but it does not exceed the optically thick regime. When scattering opacity is included, the total opacity increases at all wavelengths, such that the optical depths are just below 1 at the ring positions. This disk is not consistent with a small-grain solution ($a_{\rm max} < 300 \ \mu {\rm m}$) in the scattering case owing to its small optical depth.

The dust mass is $0.74^{+0.34}_{-0.13} \times 10^{-3} M_{\sun}$ in the nonscattering model and the same for the scattering model with large grains, consistent with the dust mass in the best-fit model in \cite{Schwarz_2020}. This dust mass is a factor of $\sim 3$ smaller than the one derived in \cite{Macias_2018}, and a factor of 2 smaller than the estimation in \cite{McClure_2016}.
The Toomre parameter $Q$ (Figure \ref{fig:toomre_stokes}) indicates that the disk is gravitationally stable, with a minimum close to the position of the outer ring at $\sim 84$ au. Based on the CO emission, \cite{Schwarz_2020} found that the region between 70 and 120 au could be gravitationally unstable, due to the total gas mass being between 0.2 and 0.3 $M_{\sun}$. However, there is no evidence of this possible instability in the dust continuum maps.
The maximum Stokes number is close to 0.1 in the inner disk and slowly decays to 0.05 at $\sim 60$ au. Then, it increases again in the outer disk. 

\subsubsection{AS 209} \label{subsec:AS209}
AS 209 is a $\sim 1$ Myr old T Tauri star with a mass of 1.2 $M_{\sun}$ \citep{Andrews_2018}.
Its surrounding disk has been previously studied with ALMA Band 6 at a high angular resolution of $0.\arcsec 037$ or $\sim 4.5$ au by \cite{Guzman_2018}, revealing seven bright rings and a central component. From our radial profiles (Figure \ref{fig:intensity_profiles}), we are able to distinguish the rings at 74.1 and 120.4 au, which are plotted together with the 15.1 au ring \citep{Guzman_2018} as vertical dashed lines in our figures as a reference. \cite{Fedele_2018} found that the deepest gap at $\sim 95$ au is consistent with the presence of a planet with a mass of 0.2 $M_{\rm Jup}$, while the inner gap at $\sim 60$ au could be generated by a planet with an upper limit mass of 0.1 $M_{\rm Jup}$. The latter is under debate, as it is not required to explain the observed structures \citep{Zhang_2018, Alarcon_2020}.

At both gaps, we found some evidence of local minima of dust mass and grain size. The dust surface density depletion is higher at 95 au than at 60 au gap location; at 95 au the dust surface density decreases by $\sim 0.6$ dex compared with the density in the outer ring at 120 au. This depletion may be larger based on the high contrasts observed in Band 6 \citep{Guzman_2018}.

The dust surface density and grain size have local maxima in the rings, which is consistent with dust trapping models \citep[e.g.,][]{Lyra_2013, Ruge_2016, Sierra_2019}, where larger grains are trapped and the dust mass is enhanced. \citet{Dullemond_2018} found that the widths of both rings are smaller than the local scale height, which also supports the idea of dust trapping in these rings.
The dust surface density decreases from $\sim 1$ g cm$^{-2}$ in the inner disk ($\sim 20$ au) to $\lesssim 0.03$ g cm$^{-2}$ in the outer disk ($\sim 130$ au), similar behavior to the dust surface density computed in \cite{Perez_2012} using particle size distributions with $p=3.0$.
The total dust mass we derive is $9.1 ^{+7.6}_{-2.7}\times 10^{-4} M_{\sun}$ in the nonscattering model and $7.5^{+7.2}_{-2.1} \times 10^{-4}$ in the scattering model with large grains. These dust masses are a factor of $\sim 2-4$ larger than the dust mass of $2.6 \times 10^{-4} M_{\sun}$ estimated from the thermal continuum at 345 GHz \citep{Andrews_2009}, the fiducial model of \cite{Fedele_2018} with $3.2 \times 10^{-4} M_{\sun}$, or polarimetric data with SPHERE \citep{Avenhaus_2018} with a dust mass of $2.32 \times 10^{-4} M_{\sun}$.

The Toomre $Q$ parameter for AS\,209 reveals that this seems to be gravitationally stable. The estimated maximum Stokes number is between $2 \times 10^{-2}$ (at $\sim 25$ au) and $3 \times 10^{-1}$ (at $\sim 100$ au), with local minima in rings and local maxima in gaps (due to dust surface density morphology). The inferred Stokes numbers in the rings are consistent with the lower limits computed in \cite{Dullemond_2018} (they assume a grain size of $200 \ \mu$m) if they are scaled to the grain size derived in this work.
\cite{Rosotti_2020} estimated the ratio between the turbulent $\alpha$ parameter and the Stokes number in rings of AS 209 and HD 163296. In the case of AS 209, they found that $\alpha/\rm{St} = 0.18 \pm 0.04 $ at 74 au and $\alpha/\rm{St} = 0.13 \pm 0.02$ at 120 au.
This means that, using our estimated maximum Stokes number, the turbulent parameters at the bright ring location are $\alpha \sim 1.3\times 10^{-2}$ at 74 au and $\sim 9.1 \times 10^{-3}$ at 120 au, which are close to the fragmentation limit computed in that work, i.e. where the dust growth is prevented owing to turbulent fragmentation.

\subsubsection{HD 163296}\label{subsec:HD163296}
HD 163296 is a Herbig Ae star with an age $\gtrsim 6$ Myr and a mass of 2.0 $M_{\sun}$ \citep{Andrews_2018}. 
Its disk has been previously studied with ALMA Band 6 at high angular resolution \citep[42 mas or 4.2 au;][]{Isella_2018}, revealing four bright rings that were already identified by \cite{Isella_2016}. From our radial profiles, we are able to distinguish the rings at 14.4, 67.0, and 100.0 au, which are plotted as vertical dashed lines in our figures as a reference.

At the gap and ring positions, we obtained local minima and maxima in the dust surface density (Figure \ref{fig:Results_sigma}), respectively. However, the maximum grain size only has a maxima in the ring at 14.4 and 100 au. The ring at 66 au does not show strong evidence of dust trapping at this resolution. This can also be seen in the spectral index profiles (Figure \ref{fig:spectral_indices}), where the minimum at 100 au is deeper compared with that at 66 au. This is consistent with the results from \cite{Dullemond_2018}, where the resolved ring width at 66 au is larger than the local scale height, disfavoring the dust trapping scenario. This is not the case for the ring at 99 au, which has a resolved ring with smaller than the local scale height. Then, the origin of the dust ring at 66 au seems to be different from that of the outer ring at 99 au. In particular, \cite{Zhang_2020} found that the midplane CO snowline is located at 65 au. Thus, the origin of the dust ring at this radius seems to be a traffic jam caused by the CO depletion, where a continuum ring is expected owing to the sintering effect at a snowline \citep[e.g.][]{Okuzumi_2016}.

The estimated dust mass is $0.90^{+0.79}_{-0.22} \times 10^{-3} M_{\sun}$ in the nonscattering model, and $0.83^{+1.35}_{-0.18} \times 10^{-3} M_{\sun}$ in the scattering model with large grains. Both dust masses are a factor of 2 smaller than that estimated using ALMA and SPHERE observations in \cite{Muro-Arena_2018}, but it is within the range (if the dust gas is 100 times smaller than the gas mass) estimated by the nonresolved observations of \cite{Isella_2007}, and \cite{Tilling_2012}, where the expected dust mass lies within $0.5 \times 10^{-3} M_{\sun}$ and $0.9 \times 10^{-3} M_{\sun}$. \cite{Booth_2019} computed a total gas mass of $0.31 \ M_{\odot}$, which implies a dust-to-gas mass ratio of 1/260 when compared with the dust mass from \cite{Isella_2007}. 
The emission of HD 163296 in all bands is optically thin, except the innermost ring at Band 6, where the optical depth is around 1 (Figure \ref{fig:optical_depth}). If scattering is included, the total opacity is just below 1 in the outer rings at 66 and 100 au (Figure \ref{fig:optical_depth_Scat}). Note that the disk is optically thin beyond $\sim 40$ au even when dust scattering opacity is included, which is consistent with the discussion of \cite{Zhu_2019} based on the nontotal extinction of the CO emission in the rings.

The Toomre parameter $Q$ (Figure \ref{fig:toomre_stokes}) indicates that the entire disk is gravitationally stable with local minima at the ring position. The Toomre parameter from \cite{Booth_2019} is $\sim 6$ at 110, close to the $\sim 4$ estimated in this work at the same position.
The maximum Stokes number lies within $\sim 10^{-2}$ in the inner disk and reaches a value of 0.3 in the outer disks, with local minima at the rings. Similarly to AS 209, the Stokes numbers in the rings are consistent with the lower limits in \cite{Dullemond_2018}.
According to the \cite{Rosotti_2020} model, the ratio between the turbulent parameter and the maximum Stokes number at 67 and 100 au is $0.23 \pm 0.03$ and $0.04 \pm 0.01$, respectively. 
Using our estimates for the maximum Stokes number, this means that the turbulent parameters at these ring radii are $\alpha \sim 9.2 \times 10^{-3}$ and  $\sim 2.8\times 10^{-3} $, respectively, consistent with constraints from line observations \citep[e.g.,][]{Flaherty_2017}. Similar to AS 209, these values are close to the turbulent fragmentation limit, which could be preventing dust growth at these positions, where, according to the dust trapping models, the effects of radial drift become inefficient \citep[e.g.,][]{Whipple_1972}.

\subsubsection{MWC 480}\label{subsec:MWC480}
MWC 480 is a 7 Myr old Herbig Ae star with a mass of 2.1 $M_{\sun}$ \citep{Simon_2019}. 
The MWC\,480 disk was studied with ALMA at Band 6 with high angular resolution (0.\arcsec 14 , $\sim 22$ au) by \cite{Long_2018} and \cite{Liu_2019b}, finding a gap centered at $\sim 74$ au and a ring at $\sim 98$ au. Additionally, \cite{Long_2018} (using the residual/uv data) and \cite{Law_2020_rad} (studying the radial profiles from the images) inferred a dark and bright ring at 149 and 165 au, respectively. The bright rings at 98 and 166 au can be identified in our data, and vertical dashed lines are plotted in our figures as a reference.

\cite{Liu_2019b} found that the observed substructure is consistent with an embedded planet with a mass of 2.3 $M_{\rm Jup}$ at a radius of 78 au. At this radius, we found a minimum in the dust surface density (Figure \ref{fig:Results_sigma}) and no clear evidence of a local minimum for the maximum grain size (Figure \ref{fig:Results_amax}). However, this minimum cannot be dismissed, because our angular resolution is larger than the width of the gap resolved with higher angular resolution. 
Similarly, at the ring position we found a maximum in the dust surface density, but the maximum grain size does not have a local maximum. In general, the dust surface density decays from $\sim 1$ g cm$^{-2}$ in the inner disk ($\sim 20$ au) to $\sim 0.03$ g cm$^{-2}$ in the outer disk ($\sim 120$ au), while the maximum grain size is $\sim 4$ mm in the inner disk and $\sim 2$ mm in the outer disk.

The estimated dust mass is $0.97 ^{+0.16}_{-0.18} \times 10^{-3} M_{\sun}$ in the nonscattering model, and $1.19 ^{+1.77}_{-0.29} \times 10^{-3} M_{\sun}$ in the scattering model with large grains. The dust mass in the nonscattering model is within a factor of $\sim 1.8-2.6$ smaller than that found in \cite{Pietu_2006} ($\sim 3.2 \times 10^{-3} M_{\sun}$), \cite{Guilloteau_2011} ($\sim 2.3 \times 10^{-3} M_{\sun}$), and \cite{Liu_2019b} ($1.6 ^{+0.5}_{-0.4} \times 10^{-3} M_{\sun}$). The dust mass with scattering is within a factor of $\sim 1.3-2$ smaller compared with the same references.
The emission in Band 3 appears to be optically thin, while for Band 6 the emission is optically thin outside of $\sim 35$ au. However, when scattering opacity is included, the total optical depth at Band 6 is optically thick within $\lesssim 60$ au.
The Toomre $Q$ parameter is larger than 2 at all disk radii, such that no evidence of gravitational instability is found. Finally, the maximum Stokes number is $\sim 9 \times 10^{-3}$ in the inner disks and increases to $\sim 0.1$ in the outer disk.

\section{Summary and Conclusions}\label{sec:conclusions}

In this paper we performed a multiwavelength analysis of the disks around IM\,Lup, GM\,Aur, AS\,209, HD\,163296, and MWC\,480 using ALMA observations at Bands 6 and 3 (and archival data for AS\,209 in Band 7 and GM\,Aur in Bands 4 and 7). The Band 6 observations consist of two intraband data sets at a central frequency of $226$ and $257$ GHz (1.33 and 1.17 mm), while the two intraband data sets at Band 3 with central frequencies of $94$ and $106$ GHz (3.20 and 2.84 mm) are merged to obtain an image with a higher S/N at a central frequency of $100$ GHz (3.0 mm). 
These observations provide at least three wavelengths that are used to fit the spatially resolved continuum spectrum from the intensity radial profiles. This fitting is performed by modeling the (sub)millimeter spectrum using the radiative transfer equation for two cases: when scattering is taken into account and when it is neglected. When scattering is considered, two possible solutions are consistent with observations in the inner disks of IM\,Lup ($\lesssim 60$ au), HD\,163296 ($\lesssim 40$ au), and MWC\,480 ($\lesssim 60$ au). The two solutions are characterized by a large-grain size regime ($a_{\rm max} > 300 \ \mu$m) and small-grain size regime ($a_{\rm max} < 300 \ \mu$m).

The dust temperature is fixed by the midplane temperature of \cite{Zhang_2020}, and a large parameter space of possible dust surface densities and grain sizes is explored. From this procedure, we constrain the radial profiles of dust surface density, maximum grain size, and optical depths at all observed wavelengths, which are able to explain the spectral properties of each disk.
The dust opacity properties are computed using a slope of the particle size distribution of $p=2.5$, which represents the case where the maximum grain size is limited by drift \citep{Birnstiel_2012}. This slope also gives a lower limit to the maximum grain size.
Once these properties are computed, radial profiles of the Toomre $Q$ parameter and the maximum Stokes number are estimated for the five disks. 
Our main conclusions can be summarized as follows.
\begin{enumerate}
    \item Millimeter grain sizes are inferred in the disks around IM\,Lup, GM\,Aur, AS\,209, HD\,163296, and MWC\,480 in the nonscattering model and scattering model in the large-grain size regime. The maximum grain size radial profiles have a negative slope from the inner disk (where the grain sizes are close to 1 cm) to the outer disk (where the grain sizes are slightly below 1 mm). The maximum grain size locally peaks in most of the known rings in these disks. While in a few cases we do not find evidence for large grains in the ring locations, but this could be due to insufficient angular resolution.
    In these models, the dust surface density decreases with disk radius in all the disks in our sample, but with local maxima that coincide with the ring positions. Then, both the dust mass and the grain size are enhanced within most of the rings, which is consistent with models where pressure maxima acts as dust size differential traps (large grains are more concentrated around the pressure maxima than small grains). 
    
    \item The inner disks ($\lesssim 20$ au) in IM\,Lup, HD\,163296, and MWC\,480 are found to be optically thick at Band 6, even if scattering opacity is not included. When scattering is taken into account, the total optical depths increase by a factor of 10; this occurs because the albedo of millimeter grain sizes is large at radio frequencies.
    
    \item Grains of a few hundred micrometers in size are consistent with the emission from the inner disks of IM\,Lup, HD\,163296, and MWC\,480, when scattering is taken into account. This alternative solution occurs because scattering modifies the optically thick spectral indices and degeneracy arises. The disks around GM\,Aur and AS\,209 are not consistent with some-hundreds-micrometer sizes because they are optically thin, even in their inner disks. This particular solution has two problems: it requires a large amount of dust in small grains to reach the level of emission at all wavelengths, making the inner disks gravitationally unstable, and it is unable to reproduce the observed emission at all wavelengths for the outer disks. Additionally, the expected total flux of IM\,Lup and HD\,163296 at $\lambda = 7$ mm from the few-hundred-micrometer grains model is low compared with their observed flux at this wavelength, while it is consistent with the expected flux from millimeter grain sizes.
    
    \item Our results strengthen the idea that IM\,Lup (which presents spiral arm structures) is a gravitationally unstable disk, as our estimated Toomre $Q$ parameter is lower than 2 outside of $\sim 15$ au. In contrast, the Toomre $Q$ parameter of GM\,Aur, AS\,209, HD\,163296, and MWC\,480 is larger than 2 at all disk radii.
    
    \item We estimate maximum Stokes numbers below 1 for all the disks. In general, the Stokes numbers in the four of disks increases with the disk radius (except IM Lup), with local minima at the ring positions, where the dust surface density is a maximum. The estimated maximum Stokes numbers, combined with the constraints on the $\alpha/{\rm St}$ ratio of \cite{Rosotti_2020}, suggest that dust growth could be limited by turbulent fragmentation in the rings of AS\,209 at 74 and 120 au and in the rings of HD\,163296 at 67 au and 100 au. 
    
    \item Scattering is an important component of opacity in protoplanetary disks. The models where scattering is taken into account provide a more detailed approximation to the radiative transfer problem in disks. In many cases, the scattering model reduces to the nonscattering model because the emission is optically thin, where the scattering effects can be neglected. However, the main differences are found in the inner disks, where the optical depth can be larger than 1.
\end{enumerate}

Observations at high angular resolution and sensitivity at multiple wavelengths are needed to better constrain the dust properties in protoplanetary disks. The degeneracy from the scattering models could be avoided if the (sub)millimeter SED is better sampled. For example, observations at longer wavelengths (e.g. 7 mm) can help to reach the midplane in the inner disk, where the ALMA wavelengths could not be tracing the dust properties. The solution that requires few hundreds of micrometer sizes in the inner disks could be rejected or strengthened with such observations. 
Also, higher angular resolution observations can be used to infer higher contrasts between the dust properties in rings and gaps, and this will help us to understand the origin of these substructures.
Finally, further modeling that incorporates polarization data would help to simultaneously explain the polarization patterns and spectral index, such that the solutions with small and large grains in the scattering models could be consistent.

\acknowledgments

We are grateful to the anonymous referee for useful comments that helped to clarify some aspects of the paper.
This paper makes use of the following ALMA data: ADS/JAO.ALMA\#2015.1.00678.S, ADS/JAO.ALMA\#2018.1.01055.L, \\ ADS/JAO.ALMA\#2017.1.01151.S.
ALMA is a partnership of ESO (representing its member states), NSF (USA) and NINS (Japan), together with NRC (Canada), MOST and ASIAA (Taiwan), and KASI (Republic of Korea), in cooperation with the Republic of Chile. The Joint ALMA Observatory is operated by ESO, AUI/NRAO and NAOJ. The National Radio Astronomy Observatory is a facility of the National Science Foundation operated under cooperative agreement by Associated Universities, Inc.

A.S.\ and L.M.P.\ acknowledge support from ANID/CONICYT Programa de Astronom\'ia Fondo ALMA-CONICYT 2018 31180052.
A.S.\ thanks Tania Serrano, Enrique Macias, Carlos Carrasco, and Osmar Guerra for helpful discussions.
L.M.P.\ acknowledges support from ANID project Basal AFB-170002 and from ANID FONDECYT Iniciaci\'on project No. 11181068. 
C.J.L. acknowledges funding from the National Science Foundation Graduate Research Fellowship under grant DGE1745303.
V.V.G. acknowledges support from FONDECYT Iniciaci\'on 11180904 and ANID project Basal AFB-170002.
K.Z. acknowledges the support of the Office of the Vice Chancellor for Research and Graduate Education at the University of Wisconsin – Madison with funding from the Wisconsin Alumni Research Foundation and support of the support of NASA through Hubble Fellowship grant HST-HF2-51401.001 awarded by the Space Telescope Science Institute, which is operated by the Association of Universities for Research in Astronomy, Inc., for NASA, under contract NAS5-26555. 
A.D.B. acknowledges support from NSF AAG grant No. 1907653.
C.W. acknowledges financial support from the University of Leeds, STFC and UKRI (grant numbers ST/R000549/1, ST/T000287/1, MR/T040726/1).
F.L. and R.T. acknowledge support from the Smithsonian Institution as a Submillimeter Array (SMA) Fellow.
F.M. acknowledges support from ANR of France under contract ANR-16-CE31-0013 (Planet-Forming-Disks)  and ANR-15-IDEX-02 (through CDP ``Origins of Life"). 
G.C. is supported by NAOJ ALMA  Scientific Research Grant Code 2019-13B.
I.C. was supported by NASA through the NASA Hubble Fellowship grant HST-HF2-51405.001-A awarded by the Space Telescope Science Institute, which is operated by the Association of Universities for Research in Astronomy, Inc., for NASA, under contract NAS5-26555.
J.B. acknowledges support by NASA through the NASA Hubble Fellowship grant No. HST-HF2-51427.001-A awarded  by  the  Space  Telescope  Science  Institute,  which  is  operated  by  the  Association  of  Universities  for  Research  in  Astronomy, Incorporated, under NASA contract NAS5-26555.
S.M.A. and J.H. acknowledge funding support from the National Aeronautics and Space Administration under grant No. 17-XRP17 2-0012 issued through the Exoplanets Research Program.  J.H. acknowledges support for this work provided by NASA through the NASA Hubble Fellowship grant No. HST-HF2-51460.001-A awarded by the Space Telescope Science Institute, which is operated by the Association of Universities for Research in Astronomy, Inc., for NASA, under contract NAS5-26555.
J.B.B. acknowledges support from NASA through the NASA Hubble Fellowship grant No. HST-HF2-51429.001-A, awarded by the Space Telescope Science Institute, which is operated by the Association of Universities for Research in Astronomy, Inc., for NASA, under contract NAS5-26555.
J.D.I. acknowledges support from the Science and Technology Facilities Council of the United Kingdom (STFC) under ST/T000287/1.
K.I.\"O. acknowledges support from the Simons Foundation (SCOL No. 321183) and an NSF AAG grant (No. 1907653). 
R.L.G. acknowledges support from a CNES fellowship grant.
T.T. is supported by JSPS KAKENHI grant Nos. JP17K14244 and JP20K04017.
Y.Y. is supported by IGPEES, WINGS Program, the University of Tokyo.
Y.A. acknowledges support by NAOJ ALMA Scientific Research Grant Code 2019-13B, and Grant-in-Aid for Scientific Research 18H05222 and 20H05847.
Y.L. acknowledges the financial support by the Natural Science Foundation of China (Grant No. 11973090).

\software{Astropy \citep{astropy:2013, astropy:2018}, CASA \citep{McMullin_2007}, Frankenstein \citep{Jennings_2020}, Matplotlib \citep{Matplotlib_2007}, Numpy \citep{Numpy_2020}}

\appendix

\section{Imaging and radial profiles}\label{app:RadProfiles}
To analyze how the radial intensity profiles in Figure \ref{fig:intensity_profiles} depend on how we image the visibilities, we compute the radial profiles in three different ways and then compare them. Here we define $\theta_{B3}$ as the beam obtained from imaging the Band 3 dataset with robust = -2 (uniform weighting), which always corresponds to the largest beam of our multiwavelength data set. The different ways one can derive the intensity profiles are as follows:

\begin{enumerate}
    \item Imaging of all datasets with robust = -2 (uniform weighting), then convolving to a common resolution given by $\theta_{B3}$, and, finally, azimuthally averaging these common resolution images to obtain the intensity radial profiles.
    \item Imaging the Band 6 datasets with robust = 0, then convolving to a common resolution given by $\theta_{B3}$, and, finally, azimuthally averaging these common resolution images to obtain the intensity radial profiles.
    \item Fitting the visibilities with a nonparametric model to derive a radial profile of the emission, which we convolve to the common $\theta_{B3}$ resolution.
\end{enumerate}

The radial profiles used in this work correspond to the first method proposed. Due to the uncertainty in using a robust parameter of -2, and because the extended emission could be filtered out in Band 6, we adopt a second method: imaging the Band 6 data with a robust parameter of 0 (still higher angular resolution than $\theta_{B3}$), and then convolving these images to the same angular resolution given by $\theta_{B3}$. Azimuthally averaged intensity profiles are also computed over the image plane in this case.
In the third case, we use the Frankenstein code \citep{Jennings_2020} to fit the visibility data (assuming that the disks are axisymmetric and using the standard hyperparameters $\alpha=1.05$, $w_{\rm smooth} = 10^{-4}$) and obtain deconvolved intensity profiles that did not go through the imaging process of the first or second method. These intensity profiles are then convolved to the same angular resolution $\theta_{B3}$.

\begin{figure}
    \centering
    \includegraphics[width=\textwidth]{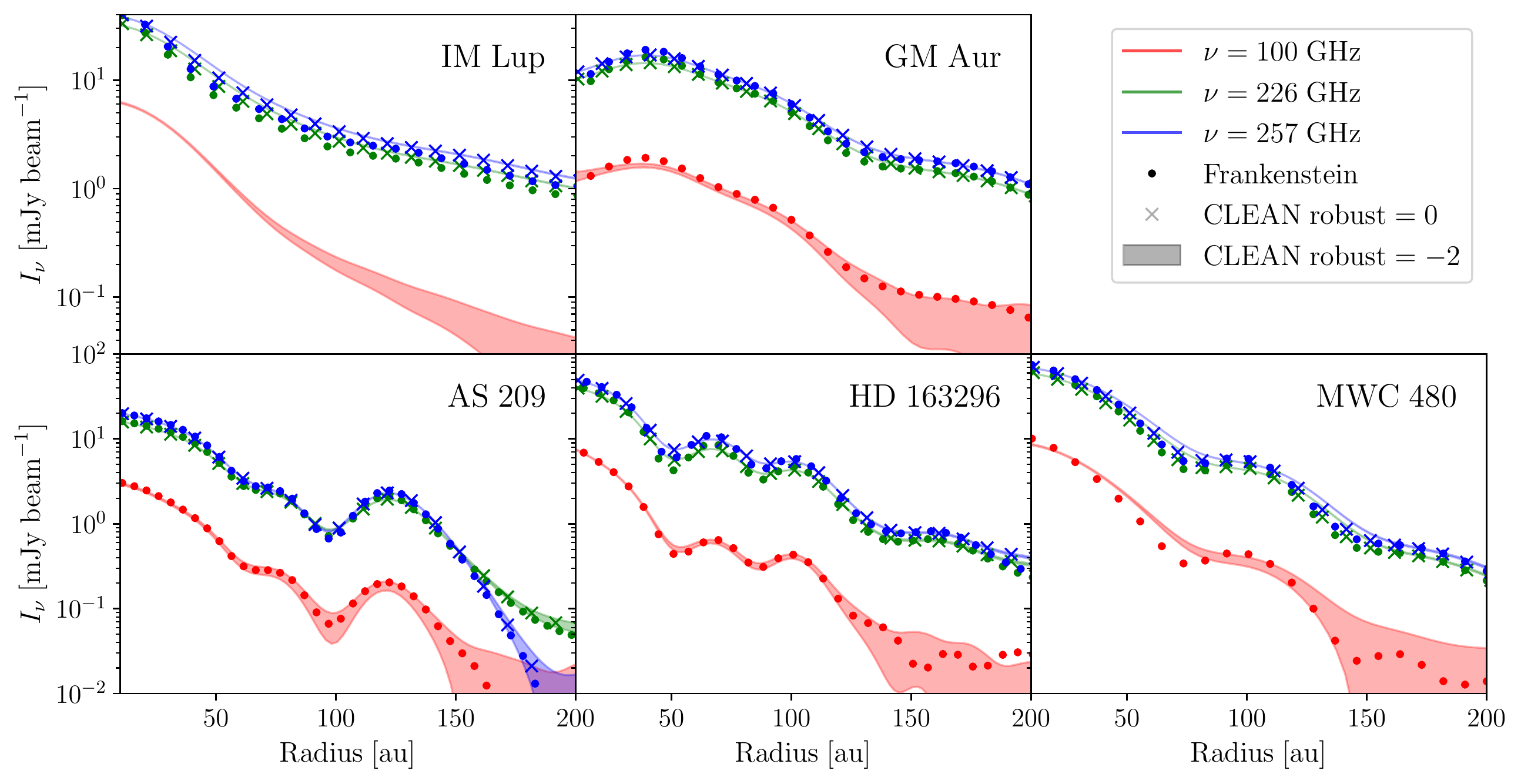}
    \caption{Convolved intensity profiles at B3 (red), B6(I) (green), and B6(II) (blue) of each disk (see legend in the upper right corner of each panel), and three different methods: fitting the visibilities (circles), robust=0 and convolution (crosses), and robust=-2 and convolution (shaded area).}
    \label{fig:RadProfiles}
\end{figure}

Figure (\ref{fig:RadProfiles}) shows the intensity profiles derived from the three methods for all disks. The shaded areas correspond to the radial profiles used in this work (method 1, robust = -2). The crosses correspond to the radial profile from method 2 with a robust parameter of 0 (the error area is not included because they overlap to those in method 1). The circles correspond to the radial profiles from fitting the visibilities (method 3). All these profiles agree, except IM Lup Band 3, where we were not able to use Frankenstein to model the visibilities, and for that reason it is not shown.

\begin{figure}
    \centering
    \includegraphics[width=\textwidth]{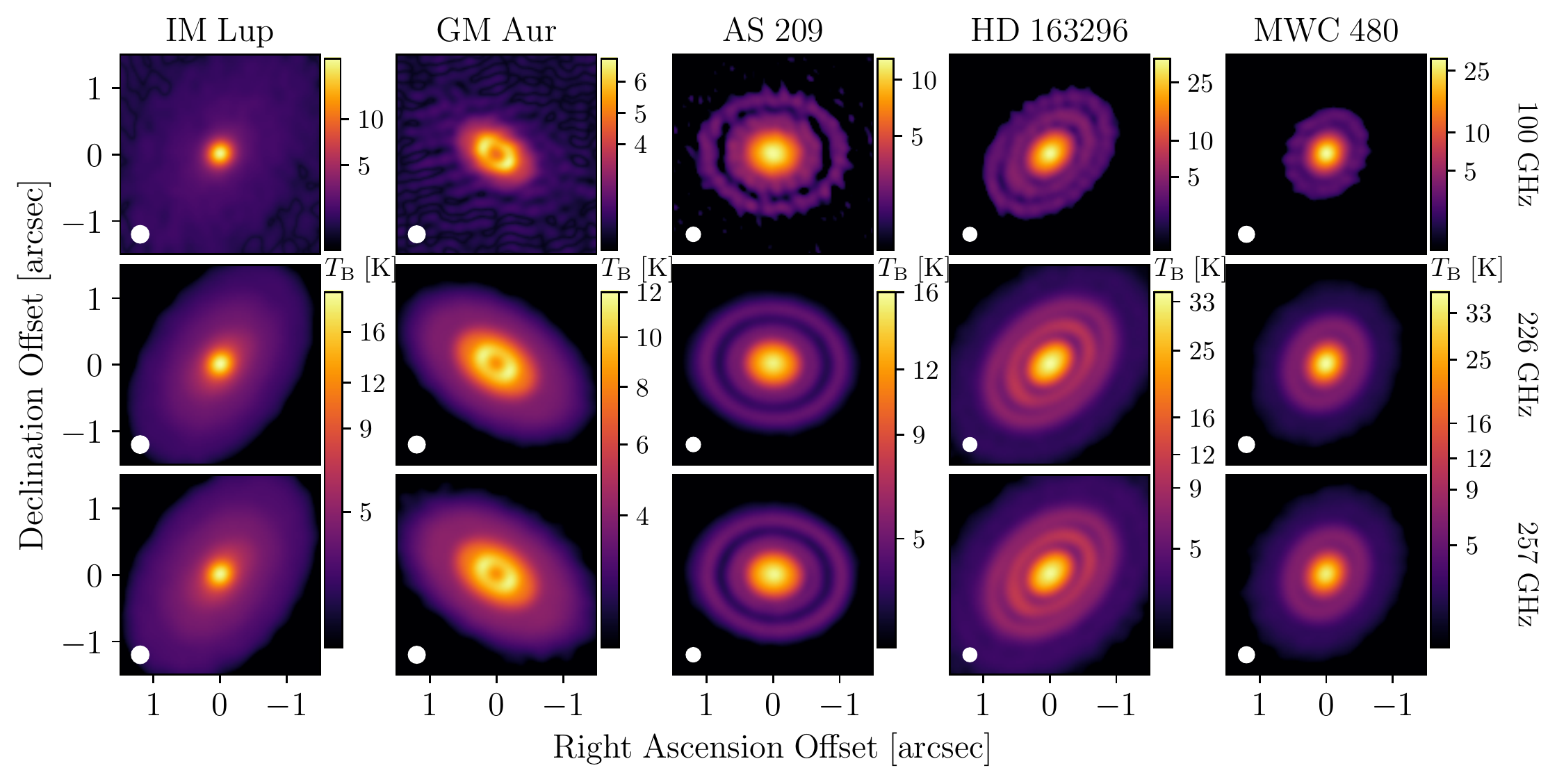}
    \caption{Brightness temperature images of the disks obtained from CLEAN with a robust parameter = -2. The top, middle, and bottom panels are observations at 100, 226, and 257 GHz, respectively. }
    \label{fig:FinalImages_clean}
\end{figure}

\begin{figure}
    \centering
    \includegraphics[width=\textwidth]{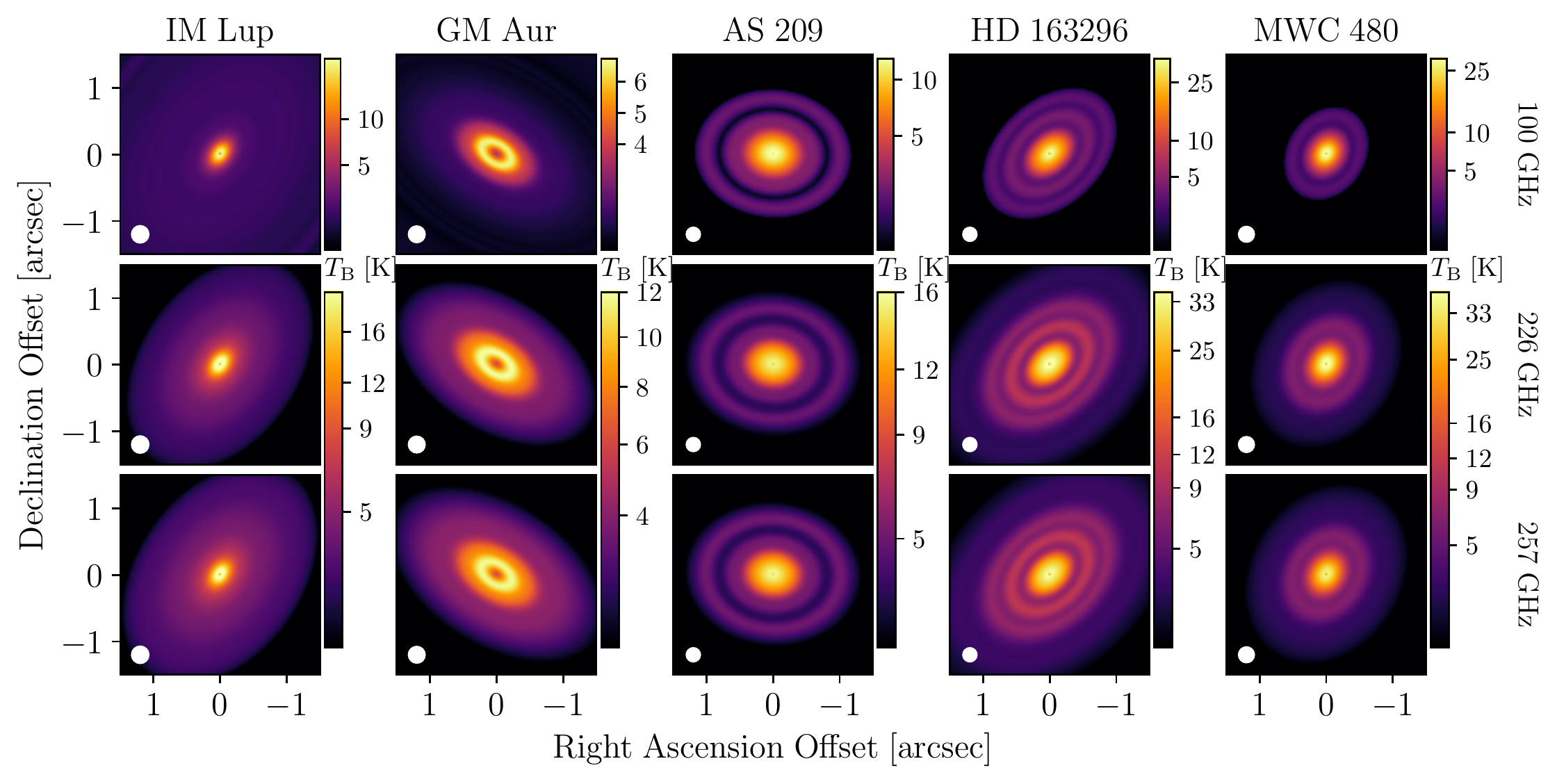}
    \caption{Brightness temperature images of the disk obtained from Frankenstein. The top, middle, and bottom panels are the observations at 100, 226, and 257 GHz, respectively.}
    \label{fig:FinalImages_frank}
\end{figure}

Figures \ref{fig:FinalImages_clean} and \ref{fig:FinalImages_frank} show the brightness temperature images of the final products from CLEAN (method 1) and Frankenstein (method 3), respectively. The maps from CLEAN are noisier than those shown in Figure \ref{fig:continuum_MAPS}, since they are computed using a smaller robust parameter. A robust = -2.0 improves the angular resolution but leads to reduced S/N. The latter is improved by azimuthally averaging the data, as the uncertainty in the radial profiles decreases by a factor of $\sqrt{n}$, where $n$ is the number of beams within a ring of the disk.

The images from Frankenstein are similar to those obtained from CLEAN, and the radial profiles match (Figure \ref{fig:RadProfiles}), so then, we confirm that false structures are not introduced when imaging the data using a robust parameter = -2.

Recently, the DSHARP observations of IM Lup, AS 209, and HD 163296 were modeled in \cite{Jennings_2021} using Frankenstein. They found deeper gaps and brighter rings in the inner region ($\sim$ 10 au) of AS 209 and HD 163296 compared with the previous constraints in \cite{Huang_2018}. The DSHARP observations are able to resolve these small substructures (compared with this work), as they probe baselines upon 10 M$\lambda$, while the observations presented here cover baselines $\lesssim 2$ M$\lambda$. Our goal, rather than finding small-scale substructure \citep[with all the continuum  substructures presented in][]{Law_2020_rad}, is to double-check that the radial profiles from CLEAN can also be recovered by an alternative methodology.

\section{Degeneracy of the inferred dust properties}\label{app:degeneracy}
Here we explore the degeneracy in the inferred dust parameters in the scattering and nonscattering cases.
We assume that the dust temperature is known, as we did in our analysis. Then, given a maximum grain size ($a_{\rm max}$) and dust surface density ($\Sigma_{\rm d}$), the intensities ($I_{\nu}$) at many wavelengths can be computed (Equation \ref{eq:Intensity_no_scat} or \ref{eq:Intensity_scatI}),

\begin{equation}
(a_{\rm max}, \Sigma_{\rm d}) \Rightarrow (I_{\nu_1},I_{\nu_2}, \dots , I_{\nu_n}).
\end{equation}

We want to know whether there is a different set of parameters ($a_{\rm max}^{\prime}$, $\Sigma_{\rm d}^{\prime}$) such that they can have a similar continuum spectrum, i.e.,

\begin{equation}
(a_{\rm max}^{\prime}, \Sigma_{\rm d}^{\prime}) \Rightarrow (I_{\nu_1}^{\prime},I_{\nu_2}^{\prime}, \dots , I_{\nu_n}^{\prime}) \approx (I_{\nu_1},I_{\nu_2}, \dots , I_{\nu_n}).
\end{equation}

We define the probability of inferring ($a_{\rm max}^{\prime}$, $\Sigma_{\rm d}^{\prime}$) given the ``real" model ($a_{\rm max}^{\rm r}$, $\Sigma_{\rm d}^{\rm r}$) as
\begin{equation}
p(a_{\rm max}^{\prime}, \Sigma_{\rm d}^{\prime} \vert a_{\rm max}^{\rm r}, \Sigma_{\rm d}^{\rm r}) = \exp (-\chi^2/2),
\label{eq:prob_amax_sigma}
\end{equation}
where the chi-squared statistic is given by
\begin{equation}
    \chi^2 = \sum_{\nu}\left( \frac{I_{\nu} - I_{\nu}^{\prime}}{\epsilon_{\nu}}\right)^2,
\end{equation}
where  $\epsilon_{\nu} =0.1 I_{\nu}$ is the error in the intensities, which is assumed to be the same as the uncertainty from the flux calibration. We assume that the observed frequencies are the same as in this work: $\nu = 257, \ 226$, and  $100$ GHz. 

To explore possible degeneracies in our model, we include random fluctuations in the intensities and to look for parameters that result in an equivalent continuum spectrum.
For this, given a set of ``real" parameters ($a_{\rm max}^{\rm r}, \Sigma_{\rm d}^{\rm r}$), the intensities $I_{\nu}$ over our wavelengths are computed. Then, a random flux factor is applied independently to each wavelength, and one can find a set of inferred parameters $a_{\rm max}^{\prime}, \Sigma_{\rm d}^{\prime}$ that have a similar continuum spectrum given their probability (Equation \ref{eq:prob_amax_sigma}).

This process is repeated many times using a random flux factor that follows a Gaussian distribution centered at one with a width of 0.1 (the uncertainty in the flux calibration). When a set of parameters ($a_{\rm max}^{\prime}, \Sigma_{\rm d}^{\prime}$) with a high probability is found (we choose $p>0.9$), these properties are plotted against the real parameters ($a_{\rm max}^{\rm r}, \Sigma_{\rm d}^{\rm r}$).

\begin{figure}
    \centering
    \includegraphics[width=\textwidth]{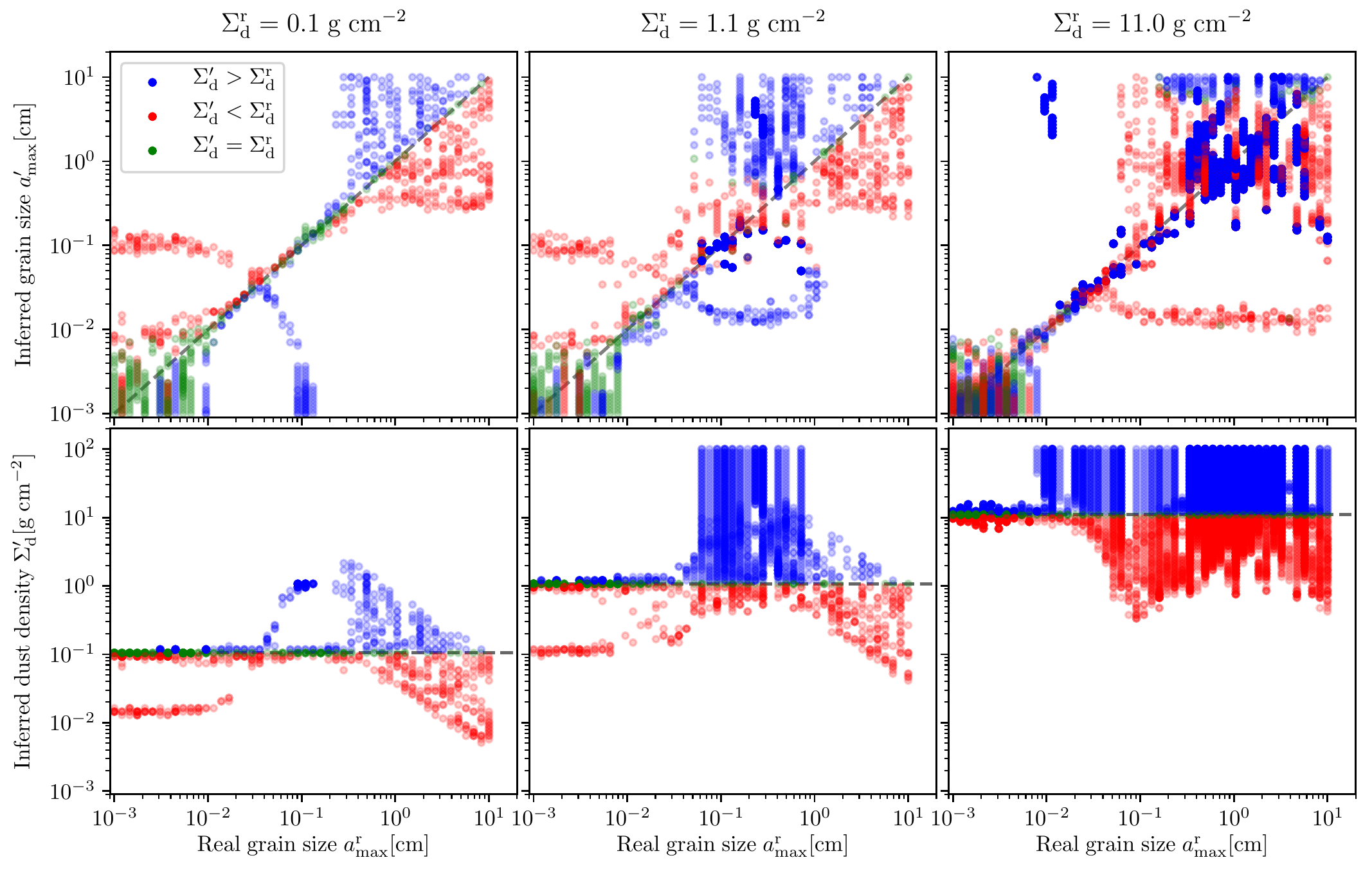}
    \caption{Maximum grain size and dust surface density degeneracy for the particular set of wavelengths observed. The left, middle, and right panels correspond to a model with a dust surface density of 0.1, 1.1, 11.0 g cm$^{-2}$, respectively. Top and bottom panels are the inferred maximum grain size and dust surface density, respectively, that have a similar SED at the three observed wavelengths (B3, B6(I), B6(II)). The blue, red, and green colors represent models where the dust surface density is overestimated, underestimated, and recovered, respectively.}
    \label{fig:Deg_ScatTrue}
\end{figure}

Figure \ref{fig:Deg_ScatTrue} shows the results for this process in a model with scattering properties. The left, middle, and right panels correspond to a model where the the real dust surface density is $\Sigma_{\rm d}^{\rm r} = 0.1, 1.1, 11.0$ g cm$^{-2}$, respectively. These values were chosen to cover different optical depth regimes.
The top panels show the inferred grain size as a function of the real grain size, and bottom panels show the inferred dust surface density as a function of the real grain size. Different colors are used to represent inferred parameters that overestimate (blue), underestimate (red), or recover (green) the real dust surface density.

In principle, the realizations shown in the top panels of Figure \ref{fig:Deg_ScatTrue} should follow a line with a slope of one, while the bottom panels should follow a horizontal line at the real dust surface density value. Although most realizations recover the original dust properties, close to this line, there are some where the inferred dust surface density and grain size differ from the real ones.

For example, in the case of $\Sigma_{\rm d}^{\rm r} = 0.1$ g cm$^{-2}$ (optically thin regime), the continuum spectrum of dust grains with a real size of $\lesssim 100\mu$m is similar to that of grains with a size of $\sim 1$ mm (red horizontal branch), but with a smaller dust density ($\Sigma_{\rm d}^{\prime} \sim 10^{-2} \rm {g \ cm}^{-2}$). This is due to the degeneracy of the opacity spectral index (middle panel of Figure 4 in \cite{Birnstiel_2018}). 
In the same way, the continuum spectrum of millimeter grains is similar to that of small grains (blue vertical branch), but with a larger inferred dust density.
The continuum spectrum of grains larger than $\sim 1$ mm is also consistent with models that overestimate the dust mass and grain size, or with models that underestimate the dust mass and grain size.

In the middle panels ($\Sigma_{\rm d}^{\rm r} = 1.1$ g cm$^{-2}$), the disks are in a transition between the optically thin and optically thick regime. The opacity of small grains ($a_{\rm max}^{\rm r} < 100 \mu$m) is very small at millimeter wavelengths, such that the emission is still in the optically thin regime, and the degeneracy is equivalent to the previous case.
The continuum spectrum of millimeter grains is similar to that of grain sizes around $\sim 200-300 \mu$m that overestimates the dust density (blue horizontal branch).
The degeneracy of grains larger than some millimeters is similar to the optically thin case, but with a significant increase of the number of models that can fit the continuum spectrum.

In the right panels ($\Sigma_{\rm d}^{\rm r} = 11$ g cm$^{-2}$), where the disks are optically thick, the degeneracy between the small grains ($a_{\rm max}^{\rm r} < 100 \mu$m) and millimeter grains disappears. However, the degeneracy between millimeter grain sizes and hundred-micrometer grains remains. This degeneracy extends to centimeter grain sizes (horizontal red branch).

In the optically thin regime (bottom left panel of Figure \ref{fig:Deg_ScatTrue}), the dust surface density could be underestimated for $\lesssim 100 \ \mu$m  and for centimeter grains. For $\gtrsim 3$ mm grains, the number of models that tends to overestimate the dust mass is larger than those that underestimate it.
In the intermediate regime (bottom middle panel of Figure \ref{fig:Deg_ScatTrue}), the mass is highly overestimated for millimeter grains, while for micrometer and centimeter grains they are equivalent to the previous optically thin case. Finally, in the optically thick regime (bottom right panel of Figure \ref{fig:Deg_ScatTrue}) the mass is well constrained for grains smaller than $\lesssim 100 \ \mu$m, but it could be overestimated or underestimated for millimeter or larger grains.
These results are for the for those wavelengths considered in this work; the addition of more observations at different wavelengths (submillimeter and centimeter) will help better constrain the dust properties in these objects.
In models where the temperature is also unknown, the degeneracy between the models will increase, unless observations at more wavelengths can help reduce the number of equivalent models.

\section{Dust temperature profiles}\label{app:Temperature}
In order to avoid degeneracy, especially in the optically thin regime, we fix the midplane temperature profiles using the models from \cite{Zhang_2020}. In their model, they iterate the density properties of a population of small grains (with sizes of $1 \ \mu$m), large grains (1 mm grains), and gas and generate dust continuum images at Band 6, which are then compared with the observed images at high resolution (Figure \ref{fig:continuum_MAPS}). From this procedure, they obtain a bidimensional structure (density and temperature) for gas and both dust populations.

We define the dust midplane temperature profiles as
\begin{equation}
    T_{\rm mid}(r) = \frac{\int \rho_{\rm d}(r,z) T(r,z)dz}{\int \rho_{\rm d}(r,z) dz},
\end{equation}
where $\rho_{\rm d}(r,z)$ is the density from the 1 mm grains and $T(r,z)$ is the dust temperature at the cylindrical coordinates ($r,z$). The integral is evaluated within an aspect ratio of $z/r =1/10$. We tested changing this aspect ratio, but the final temperature profiles are only slightly modified. This occurs because the temperature from the denser regions close to the midplane dominates the integrals.

\begin{figure}
    \centering
    \includegraphics[width=\textwidth]{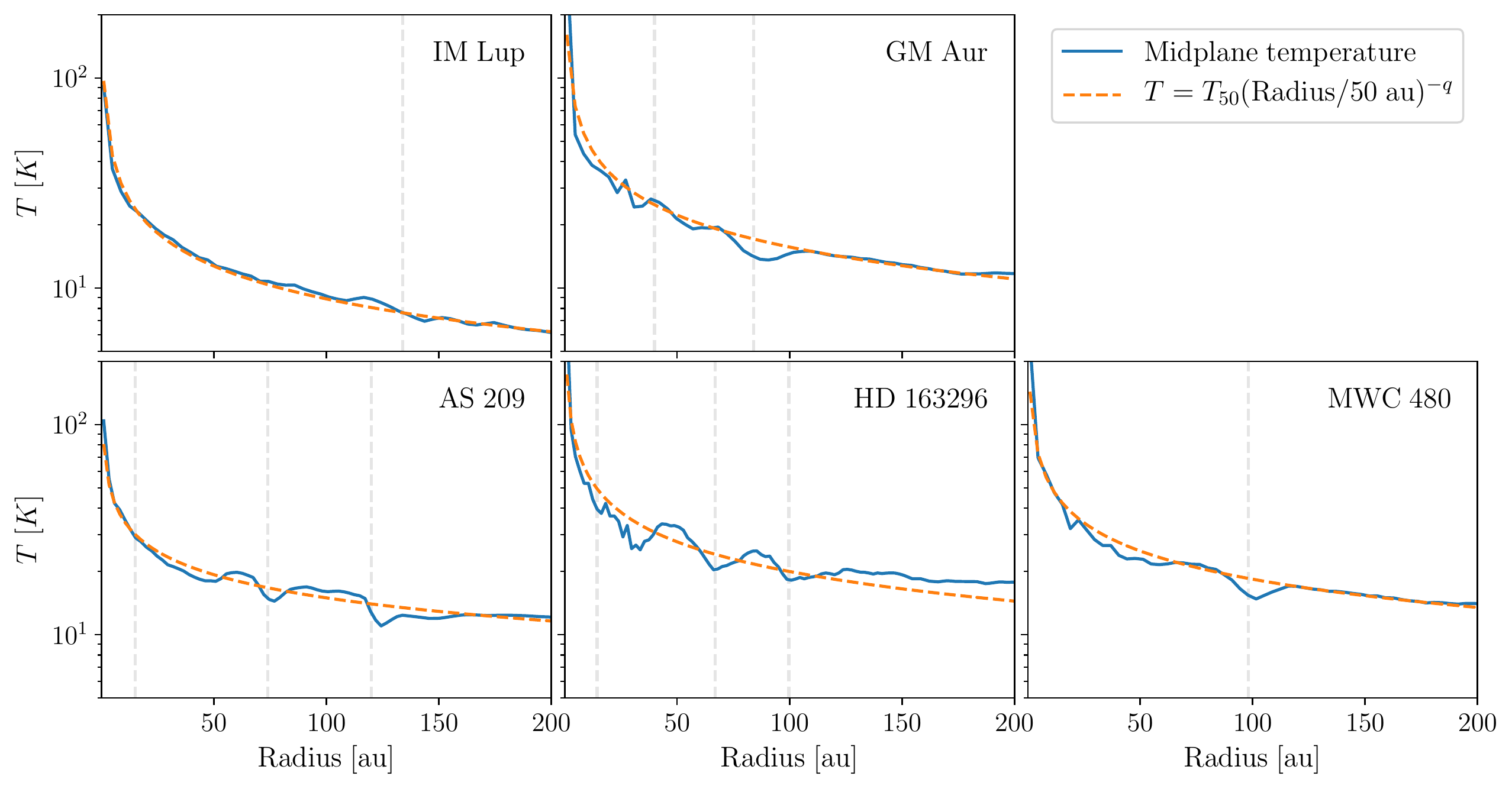}
    \caption{Dust temperature profiles for the five disks. The blue line is the midplane temperature defined by equation \ref{eq:MidTemp}, and the orange line is a power-law fit to the midplane temperature. Vertical dashed lines mark the position of bright rings in each disk.}
    \label{fig:Tmidplane}
\end{figure}

Figure \ref{fig:Tmidplane} shows the midplane temperature for the five disks (blue lines). Additionally, for reference we fitted the radial profiles to a power-law function of the radius (dashed orange line)
\begin{equation}
    T = T_{50} \left(r/50 \rm au \right)^{-q}.
    \label{eq:MidTemp}
\end{equation}
The magnitude of the temperature at 10 au and the slope of the power law from the fit are summarized in Table \ref{tab:Tmid}.

In all cases, the midplane profiles are convolved to the angular resolution of each disk and then used to fit the dust surface density and maximum grain size (Section \ref{sec:results}).

\begin{table}
    \centering
    \caption{Power-law fit of the Midplane Dust Temperature.}    
    \begin{tabular}{ccc}
        \hline
        Source & $T_{50} \ (\rm{K})$ & $q$  \\
        \hline
         IM Lup    & 12.7 $\pm$ 0.1 & 0.52 $\pm$ 0.01 \\        
         GM Aur    & 22.3 $\pm$ 0.4 & 0.51 $\pm$ 0.02 \\
         AS 209     & 19.3 $\pm$ 0.2 & 0.37 $\pm$ 0.01 \\         
         HD 163296 & 27.7 $\pm$ 1.3 & 0.46 $\pm$ 0.04 \\
         MWC 480   & 25.0 $\pm$ 0.4 & 0.45 $\pm$ 0.01 \\
         \hline
    \end{tabular}
    \label{tab:Tmid}
\end{table}

\bibliography{main}{}
\bibliographystyle{aasjournal}

\end{document}